\documentclass[12pt]{article}

 \usepackage{amssymb}   

 \usepackage{mathtools} 
 \usepackage{graphicx}
 \usepackage{cite}

\begin{document}


\begin{center}
{\LARGE\bf Topological properties of minimally doubled\\[2mm] fermions in two space-time dimensions}
\end{center}

\vspace{20pt}

\begin{center}
{\large\bf Stephan D\"urr$\,^{a,b}$}
\,\,\,and\,\,\,
{\large\bf Johannes H.\ Weber$\,^{c}$}
\\[10pt]
${}^a${\sl Department of Physics, University of Wuppertal, 42119 Wuppertal, Germany}\\
${}^b${\sl J\"ulich Supercomputing Centre, Forschungszentrum J\"ulich, 52425 J\"ulich, Germany}\\
${}^c${\sl Physics Department, Humboldt University of Berlin, D-12489 Berlin, Germany}
\end{center}

\vspace{10pt}

\begin{abstract}
The two-dimensional Schwinger model is used to explore how lattice fermion operators perceive the global topological charge $q\in\mathbb{Z}$ of a given background gauge field.
We focus on Karsten-Wilczek and Borici-Creutz fermions, which are minimally doubled, and compare them to Wilson, Brillouin, naive, staggered and Adams fermions.
For each operator the eigenvalue spectrum in a background with $q\neq0$ is determined along with the chiralities of the eigenmodes, and the spectral flow of the pertinent hermitean operator is worked out.
We find that Karsten-Wilczek and Borici-Creutz fermions perceive the global topological charge $q$ in the same way as staggered and naive fermions do.
\end{abstract}

\vspace{20pt}

\newcommand{\pad}{\partial}
\newcommand{\hqu}{\hbar}
\newcommand{\til}{\tilde}
\newcommand{\pri}{^\prime}
\renewcommand{\dag}{^\dagger}
\newcommand{\<}{\langle}
\renewcommand{\>}{\rangle}
\newcommand{\gaf}{\gamma_5}
\newcommand{\nab}{\nabla}
\newcommand{\lap}{\triangle}
\newcommand{\dal}{{\sqcap\!\!\!\!\sqcup}}
\newcommand{\trc}{\mathrm{tr}}
\newcommand{\Trc}{\mathrm{Tr}}
\newcommand{\Mpi}{M_\pi}
\newcommand{\Fpi}{F_\pi}
\newcommand{\Mka}{M_K}
\newcommand{\Fka}{F_K}
\newcommand{\Met}{M_\et}
\newcommand{\Fet}{F_\et}
\newcommand{\Mss}{M_{\bar{s}s}}
\newcommand{\Fss}{F_{\bar{s}s}}
\newcommand{\Mcc}{M_{\bar{c}c}}
\newcommand{\Fcc}{F_{\bar{c}c}}

\newcommand{\al}{\alpha}
\newcommand{\be}{\beta}
\newcommand{\ga}{\gamma}
\newcommand{\de}{\delta}
\newcommand{\ep}{\epsilon}
\newcommand{\ve}{\varepsilon}
\newcommand{\ze}{\zeta}
\newcommand{\et}{\eta}
\renewcommand{\th}{\theta}
\newcommand{\vt}{\vartheta}
\newcommand{\io}{\iota}
\newcommand{\ka}{\kappa}
\newcommand{\la}{\lambda}
\newcommand{\rh}{\rho}
\newcommand{\vr}{\varrho}
\newcommand{\si}{\sigma}
\newcommand{\ta}{\tau}
\newcommand{\ph}{\phi}
\newcommand{\vp}{\varphi}
\newcommand{\ch}{\chi}
\newcommand{\ps}{\psi}
\newcommand{\om}{\omega}

\newcommand{\qhat}{\hat{q}}
\newcommand{\khat}{\hat{k}}

\newcommand{\bdm}{\begin{displaymath}}
\newcommand{\edm}{\end{displaymath}}
\newcommand{\bea}{\begin{eqnarray}}
\newcommand{\eea}{\end{eqnarray}}
\newcommand{\beq}{\begin{equation}}
\newcommand{\eeq}{\end{equation}}

\newcommand{\mr}{\mathrm}
\newcommand{\mb}{\mathbf}
\newcommand{\ri}{\mr{i}}
\newcommand{\Nf}{N_{\!f}}
\newcommand{\Nc}{N_{ c }}
\newcommand{\Nt}{N_{ t }}
\newcommand{\Nv}{N_{ v }}
\newcommand{\Nthr}{N_\mr{thr}}
\newcommand{\Dst}{D^\mr{st}}
\newcommand{\DW}{D_\mr{W}}
\newcommand{\DB}{D_\mr{B}}
\newcommand{\DN}{D_\mr{N}}
\newcommand{\DS}{D_\mr{S}}
\newcommand{\DA}{D_\mr{A}}
\newcommand{\DKW}{D_\mr{KW}}
\newcommand{\DBC}{D_\mr{BC}}
\newcommand{\HW}{H_\mr{W}}
\newcommand{\HB}{H_\mr{B}}
\newcommand{\HN}{H_\mr{N}}
\newcommand{\HS}{H_\mr{S}}
\newcommand{\HA}{H_\mr{A}}
\newcommand{\MeV}{\,\mr{MeV}}
\newcommand{\GeV}{\,\mr{GeV}}
\newcommand{\fm}{\,\mr{fm}}
\newcommand{\MSbar}{\overline{\mr{MS}}}



\section{Introduction}


How does a given Dirac fermion operator $D$ perceive topology ?
In lattice gauge theory, this question has been asked persistently over several decades.
The answers would be phrased in the language of three ``iconic plots'':
({\it i}\,) the chiralities $\<\ps|\gaf|\ps\>$ where $\ps$ is an eigenmode of the Dirac operator,
({\it ii}\,) the crossings of the eigenvalues of the hermitean counterpart operator, and
({\it iii}\,) the fermionic topological charge $q \simeq m\,\mr{tr}(D_m^{-1}\gaf)$ versus $m$.

In this paper we aim to produce such plots for Karsten-Wilczek \cite{Karsten:1981gd,Wilczek:1987kw} and Borici-Creutz \cite{Creutz:2007af,Borici:2007kz} fermions.
These discretization schemes are in the class of minimally doubled lattice fermion actions, i.e.\ they yield two species in the continuum limit
(precisely the minimum required by the Nielsen-Ninomiya theorem \cite{Karsten:1980wd,Nielsen:1981xu,Nielsen:1980rz}) and yet maintain an exact chiral symmetry.
Some elementary properties of these formulations like the spectral range and the free field dispersion relations were worked out in Ref.~\cite{Durr:2020yqa}.
A slight disadvantage in practical terms is that the remnant chiral symmetry is tasted;
this has been discussed in detail in Refs.~\cite{Sharatchandra:1981si,KlubergStern:1983dg,Golterman:1984cy,Smit:1986fn,Blum:1996uf,Orginos:1998ue,Lee:1999zxa,Knechtli:2000ku}.

The compliance of minimally doubled fermions with the Atiyah-Singer index theorem has been discussed in Refs.~\cite{Pernici:1994yj,Tiburzi:2010bm,Creutz:2010bm}.
In our view the topological properties of KW and BC fermions are most transparent if presented alongside ``known properties'' of more mundane formulations (Wilson, Brillouin, naive, staggered and Adams fermions).
In consequence, both the implementation effort and the amount of material to be presented proliferate, and this is why the scope of this paper is limited to two space-time dimensions (``2D'').
We are optimistic that we will follow up with a paper focusing on the situation in four space-time dimensions (``4D'').

We use the quenched Schwinger \cite{Schwinger:1962tp,Lowenstein:1971fc} model as a testbed for our calculations, since the set of all $U(1)$ gauge fields on a two-dimensional torus
falls into classes labeled by a topological index $q\in\mathbb{Z}$ \cite{Ansourian:1977qe}, like in QCD, and the theory can be
simulated without topology freezing \cite{Dilger:1992yn,Dilger:1994ma,Durr:2012te,Albandea:2021lvl,Eichhorn:2021ccz}.
In 2D fermion matrices tend to be small, and their eigenvalues may be evaluated inexpensively.
We apply one step of stout-smearing with $\rh=0.25$ \cite{Morningstar:2003gk} to the ``thin link'' gauge field $U$, and evaluate the fermion operators on the resulting ``fat link'' gauge background $V$.
A preliminary account of this investigation has been given in Ref.~\cite{Durr:2021gma}.

The remainder of this article is organized as follows.
The situation for Wilson and Brillouin fermions is reviewed in Sec.~\ref{sec:wilsbril}.
The details for staggered and Adams fermions are worked out in Sec.~\ref{sec:stagadam}.
The relation between the latter two formulations is mirrored by the relation between the naive action without and with an Adams-like taste-splitting term $C_\mr{sym}$, as shown in Sec.~\ref{sec:naiv}.
In Sec.~\ref{sec:cebr} we discuss central-branch fermions and their descendants.
Armed with this insight we are in a position to analyze the situation for KW fermions in Sec.~\ref{sec:kawi} and for BC fermions in Sec.~\ref{sec:bocr}.
The lesson learned on species-lifting terms can be applied to KW and BC fermions; in Sec.~\ref{sec:KWBC} we demonstrate that this yields single-taste formulations with additive mass renormalization.
Finally, a summary is presented in Sec.~\ref{sec:conclusions}.
Our notation and the Clifford algebra conventions are specified in App.~\ref{app:notation}.
A side-by-side comparison of all fermionic charge definitions motivated by our investigations is given in App.~\ref{app:charges},
and complemented with an analytic argument in App.~\ref{app:analytic}.


\section{Wilson and Brillouin fermions\label{sec:wilsbril}}


\begin{figure}[!tb]
\includegraphics[width=0.48\textwidth]{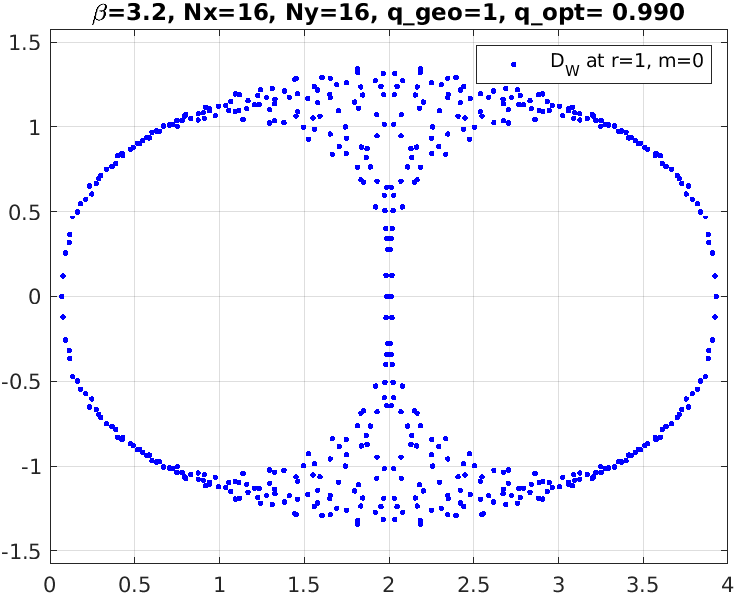}\hfill
\includegraphics[width=0.49\textwidth]{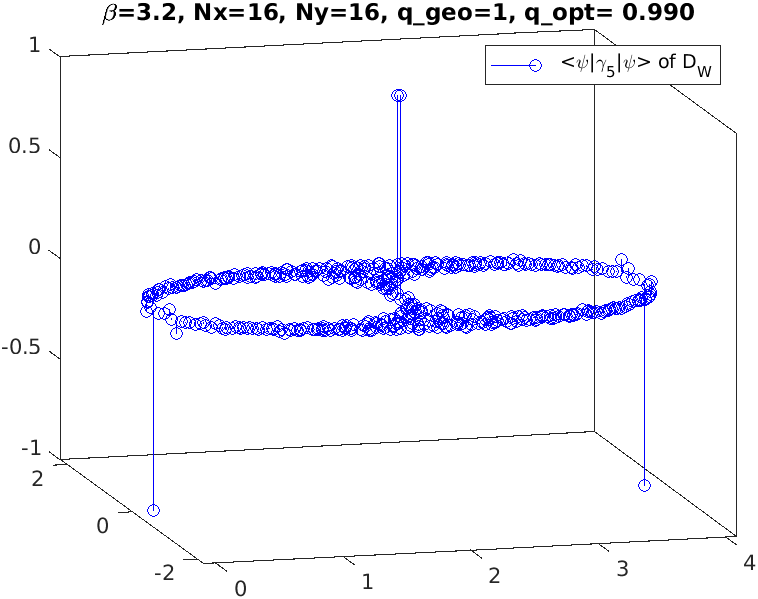}%
\vspace*{-2mm}
\caption{\label{fig:eigstem_wils}\sl
Eigenvalues of the Wilson operator on a background with $q=1$ (left), and ``needle plot''
of the $\gaf$-chiralities in the pertinent left-right-eigenvector sandwich (right).}
\vspace*{+2mm}
\includegraphics[width=0.49\textwidth]{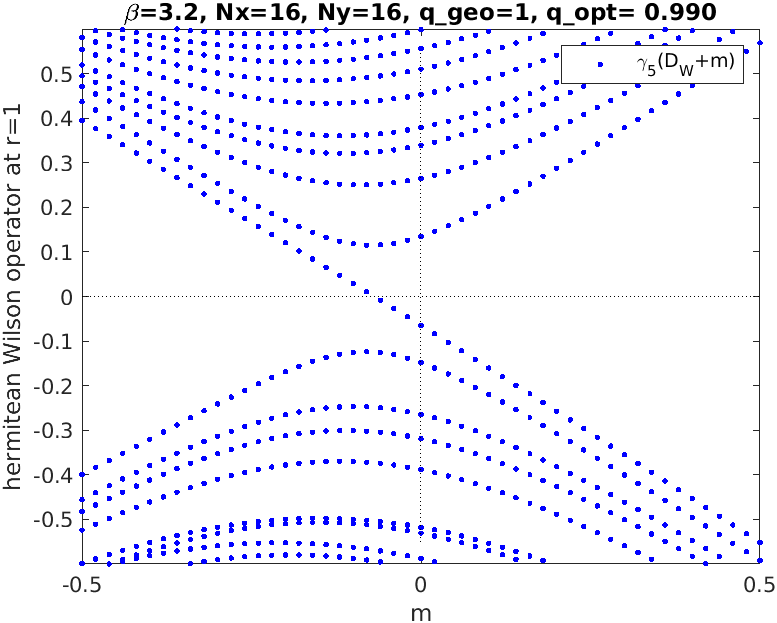}\hfill
\includegraphics[width=0.49\textwidth]{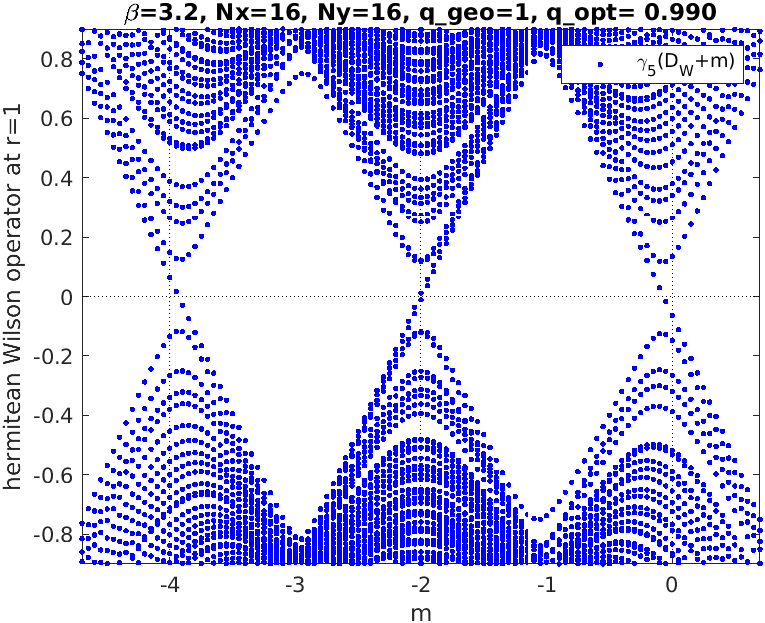}%
\vspace*{-2mm}
\caption{\label{fig:specflow_wils}\sl
Spectral flow of the Wilson operator, i.e.\ eigenvalues of $\gaf(\DW+m)$ versus $m$.
Relevant part near $m=0$ (left) and panoramic view (right).}
\end{figure}

The Wilson Dirac operator at vanishing bare mass is defined as \cite{Wilson:1974sk}
\beq
\DW(x,y)=\sum_\mu \ga_\mu \nab_\mu(x,y)
-\frac{ra}{2}\sum_\mu \lap_\mu(x,y)
\label{def_wils}
\eeq
and a glimpse at (\ref{def_naiv}) reveals that it differs from the naive Dirac operator by a hermitean, positive semi-definite term of mass dimension $5$.
Due to $\lap_\mu\dag=\lap_\mu$ and $[\lap_\mu,\gaf]=0$ the Wilson operator is $\gaf$-hermitean, i.e.\ $\gaf \DW \gaf=\DW\dag$.
An unpleasant feature is that the term $\sum_\mu \lap_\mu(x,y)$ mixes%
\footnote{Another (we think more adequate) view is that the Laplacian consists of two parts, $a^2\lap_\mu=2C_\mu-2I$, with $C_\mu$ given in (\ref{def_cmu}) and $I$ the identity,
and that $C_\mu$ (which depends on the gauge field) transforms differently under taste rotations than $I$ (present in $\lap_\mu$ and the mass term), see footnote~\ref{foot:taste} for details.}
on interacting gauge backgrounds with the identity.
Chiral symmetry is broken, and the bare mass $m$ in the massive operator $\DN+m$ is both additively and multiplicatively renormalized \cite{BOOK_MM}.
In the free-field limit the Wilson operator takes a diagonal form in momentum space
\bea
\DW(p)&=&\ri \sum_\mu \ga_\mu \frac{1}{a}\sin(ap_\mu)+\frac{r}{a}\sum_\mu \{1-\cos(ap_\mu)\}
\nonumber
\\
&=&\ri \sum_\mu \ga_\mu \bar{p}_\mu
+\frac{ra}{2} \sum_\mu \hat{p}_\mu^2
\quad \mbox{with} \quad \hat{p}_\mu=\frac{2}{a}\sin(\frac{ap_\mu}{2})
\label{momrep_wils}
\eea
which again highlights the anti-hermitean and hermitean positive semi-definite nature of the two terms, respectively.
Specifically for $r=1$ the $2^{d/2}-1$ unphysical species do not propagate into any one of the $2d$ on-axis directions \cite{BOOK_MM}.

The eigenvalues $\la_i\in\mathbb{C}$ of $\DW$ on an interacting background with topological charge $q=1$ are shown in Fig.~\ref{fig:eigstem_wils}.
The result is not far from the free-field case%
\footnote{In the free-field case the eigenvalue spectrum of $\DW$ follows from $\ga_\mu$ ($\mu=1,...,d$) having eigenvalues $\pm1$.
The eigenvalues of $\DW$ are inside an ellipse that fits into the rectangle $[0,2dr]\times[-\sqrt{d},+\sqrt{d}]$ in $d$ dimensions.}
and two depleted areas separate the physical branch at $\mr{Re}(\la)\simeq0$ from the two species at $\mr{Re}(\la)\simeq2$ and the one at $\mr{Re}(\la)\simeq4$.
The symmetry about the real axis reflects the pairing property imposed by the $\gaf$-hermiticity \cite{BOOK_MM}.
Due to the breaking of chiral symmetry the physical branch has a non-zero renormalized mass \cite{BOOK_MM},
and adding a bare mass term $m\de_{x,y}$ to (\ref{def_wils}) shifts all eigenvalues by $+m$.

We calculate both the left-eigenvector $\<\ps_i|\DW=\<\ps_i|\la_i$ and the right-eigenvector $\DW|\ps_i\>=\la_i|\ps_i\>$ for any (joint) eigenvalue $\la_i$,
with $i=1,...,2^{d/2}N_\mr{vol}$ and $N_\mr{vol}=N_1\cdot ...\cdot N_d$ the box volume in lattice units.
Since $\DW$ is non-normal%
\footnote{\label{foot:normality} An operator $A$ is normal if $[A,A\dag]=0$. In this case the row-vector $\<\ps_i|$ is the daggered version
of the column vector $|\ps_i\>$. In the event $A$ is non-normal, there is no way of obtaining $\<\ps_i|$ from a single $|\ps_j\>$.
Here, one needs to combine all column vectors $|\ps_j\>$ into a matrix, invert it, and the $i$-th row of the inverse is $\<\ps_i|$.
In the literature the latter statement features as bi-orthogonality condition $\<\ps_i|\ps_j\>=\de_{ij}$. See Ref.~\cite{Hip:2001mh} for details.}
$\<\ps_i|$ is not related to $|\ps_i\>$ by a dagger-operation.
With the left- and the right-eigenvectors in hand, one finds $\<\ps_i|\gaf|\ps_i\>$ for each $i$.
The result is plotted as a ``needle'' above the pertinent $\la_i\in\mathbb{C}$ in the right panel.
Two needles reach almost down to $-1$, indicating one species with correct chirality in the physical branch and one in the doubly-lifted branch near $\mr{Re}(\la)\simeq4$.
And two needles reach almost up to $+1$, indicating two species with opposite chirality in the singly-lifted branch near $\mr{Re}(\la)\simeq2$.

It is customary to move on to the hermitean Wilson operator $\HW=\gaf(\DW+m)$ \cite{Edwards:1998gk},
and to plot its eigenvalue spectrum (which is in $\mathbb{R}$) as a function of the bare mass $m$, see Fig.~\ref{fig:specflow_wils}.
There is one downward-crossing near $m=0$, indicating the fermionic topological charge $q_\mr{W}=+1$, in line%
\footnote{There is a difference in sign between $d=2$ and $d=4$, see the discussion in App.~\ref{app:charges}.}
with the the topological charge being $q=+1$.
However, upon adopting a panoramic view, one sees that there is no net crossing as $m$ tends from $-\infty$ to $+\infty$.

\begin{figure}[!tb]
\includegraphics[width=0.48\textwidth]{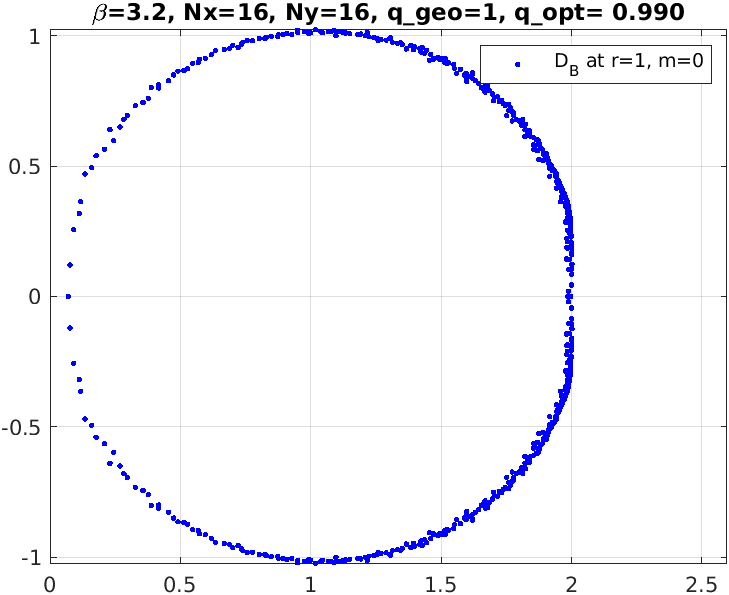}\hfill
\includegraphics[width=0.49\textwidth]{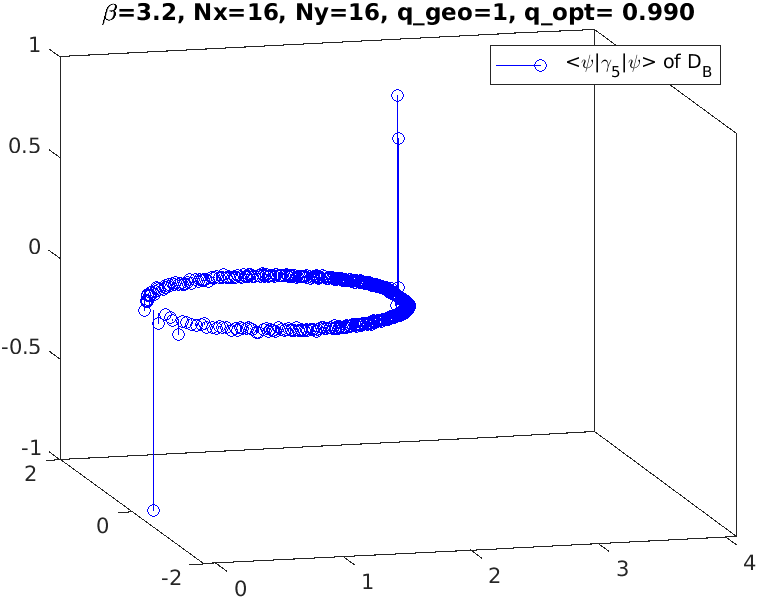}%
\vspace*{-2mm}
\caption{\label{fig:eigstem_bril}\sl
Eigenvalues of the Brillouin operator on a background with $q=1$ (left), and ``needle plot''
of the $\gaf$-chiralities in the pertinent left-right-eigenvector sandwich (right).}
\vspace*{+2mm}
\includegraphics[width=0.49\textwidth]{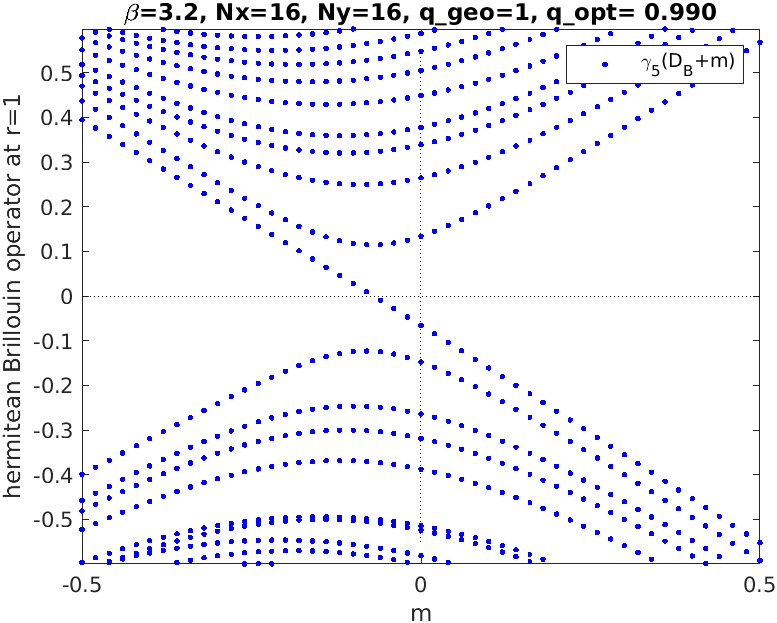}\hfill
\includegraphics[width=0.49\textwidth]{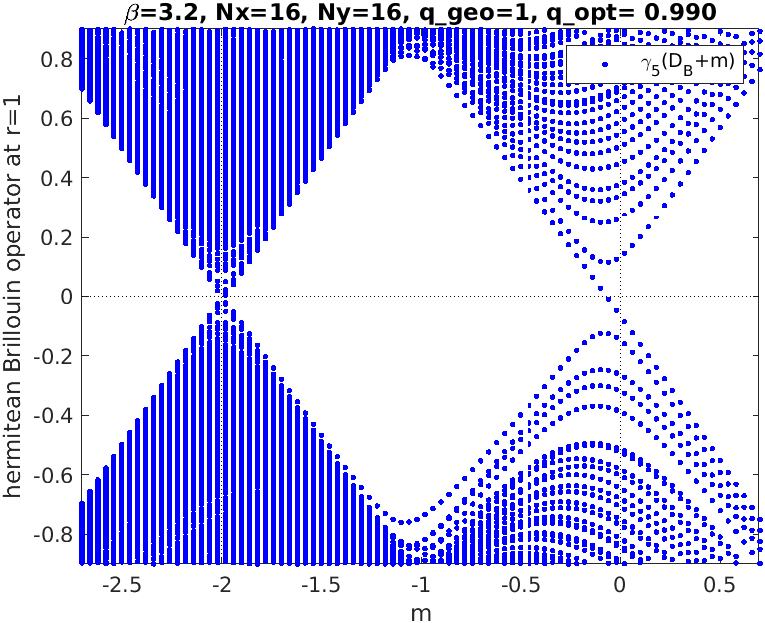}%
\vspace*{-2mm}
\caption{\label{fig:specflow_bril}\sl
Spectral flow of the Brillouin operator, i.e.\ eigenvalues of $\gaf(\DB+m)$ versus $m$.
Relevant part near $m=0$ (left) and panoramic view (right).}
\end{figure}

The Brillouin Dirac operator at zero bare mass is defined as \cite{Durr:2010ch,Durr:2017wfi}
\beq
\DB(x,y)=\sum_\mu \ga_\mu \nab_\mu^\mr{iso}(x,y)
-\frac{ra}{2}\lap^\mr{bri}(x,y)
\label{def_bril}
\eeq
where $\nab_\mu^\mr{iso}$ denotes a $2\times3^{d-1}$-point discretization of the covariant derivative, and $\lap^\mr{bri}$ denotes a $3^d$-point discretization of the gauged Laplacian.
It is conceptually a Wilson-type%
\footnote{The Brillouin fermion \cite{Durr:2010ch,Durr:2017wfi} shares this property with the closely related
hypercube fermion of Bietenholz \emph{et al.} \cite{Bietenholz:1999km,Bietenholz:2006fj} and the
chirally improved fermion of Gattringer \emph{et al.} \cite{BGR:2003glc,Gattringer:2008vj}.}
fermion, albeit with reduced breaking of the hypercubic symmetry.

The eigenvalues $\la_i\in\mathbb{C}$ of $\DB$ at $m=0$ on the same background configuration are shown in Fig.~\ref{fig:eigstem_bril}.
The main difference to the Wilson eigenvalue spectrum is that all doublers are located near $\mr{Re}(\la)\simeq2$, the physical branch near $\mr{Re}(\la)\simeq0$ contains only one species.
The additive mass shift is comparable to the Wilson case, and the symmetry about the real axis indicates that $\DB$ is $\gaf$-hermitean.
We compute the left-eigenvector $\<\ps_i|$ and the right-eigenvector $|\ps_i\>$ for each eigenvalue $\la_i$,
and plot the chirality $\<\ps_i|\gaf|\ps_i\>$ as a needle at position $\la_i\in\mathbb{C}$.
The chirality of the would-be zero-mode $\la\simeq0.06978$ is almost $-1$, while the two modes at $1.9972\pm0.0012\ri$ and the one mode at $1.9851$ mix%
\footnote{The diagonal elements in this $3\times3$ block of the chirality matrix read $0.785,0.785,1.0$, while the six off-diagonal elements in this block are not close to zero.
On the other hand, there is very little mixing between these three modes and the physical mode (out of the six extra off-diagonal elements present in the $4\times4$ matrix
the two largest in magnitude are $\pm0.00267$).}
heavily.

The eigenvalue flow of the hermitean Brillouin operator $\HB=\gaf(\DB+m)$ is shown in Fig.~\ref{fig:specflow_bril}.
There is again one eigenvalue crossing near $m=0$, indicating that the Brillouin operator finds $q_\mr{B}=+1$, too.
In the panoramic view there is no net eigenvalue crossing, though this property is not as easily seen as in the Wilson case.


\section{Staggered and Adams fermions\label{sec:stagadam}}


The Susskind (``staggered'') Dirac operator at vanishing bare mass is defined as \cite{Susskind:1976jm}
\beq
\DS(x,y)=\sum_\mu \et_\mu(x) \nab_\mu(x,y)
\label{def_stag}
\eeq
with the Kawamoto-Smit phase factors $\et_\mu(x)=(-1)^{\sum_{\nu<\mu}x_\nu}$ and $\ze_\mu(x)=(-1)^{\sum_{\nu>\mu}x_\nu}$ \cite{Kawamoto:1981hw}.
These are used to define the matrices $\Gamma_\mu(x,y)$, see (\ref{def_GA}), which act like $\ga_\mu$ in spinor space,
and the matrices $\Xi_\mu(x,y)$, see (\ref{def_XI}), which act like $\ga_\mu$ in taste%
\footnote{In the latter case the matrix $\ga_\mu$ is often denoted by $\xi_\mu$, to avoid confusion.
In such a situation $\ga_\al\otimes\xi_\be$ denotes a combined transformation in spinor and taste space,
see Refs.~\cite{Sharatchandra:1981si,KlubergStern:1983dg,Golterman:1984cy,Smit:1986fn} and Ref.~\cite{BOOK_MM}.}
space \cite{Sharatchandra:1981si,KlubergStern:1983dg,Golterman:1984cy,Smit:1986fn}.
Unlike in the Wilson case, these matrices depend on the gauge field $U$; each $\Gamma_\mu$ or $\Xi_\mu$ is a $1$-hop operator.

In the following we use the matrix $\Gamma_5(x,y)$ which implements $\gaf$ in spinor space (up to cut-off effects),
and there is a similar matrix $\Xi_5(x,y)$ which implements $\xi_5$ in taste space (up to cut-off effects).
The precise definitions are given in App.~\ref{app:notation}.
Sometimes a two-index notation is used to refer to the $\ga\otimes\xi$ decomposition, specifically $\Gamma_{50}\equiv\Gamma_5\otimes1$ and $\Gamma_{05}\equiv1\otimes\Xi_5$.
In this approach $\Gamma_5$ and $\Xi_5$ are extended ultra-local operators ($d$-hop operators in $d$ dimensions).

Furthermore, there is the $0$-hop operator $\ep(x,y)=(-1)^{\sum_\mu x_\mu}\de_{x,y}$, with the representation $\ep\doteq\gaf\otimes\xi_5$, and it is sometimes denoted $\Gamma_{55}$.
Unlike $\Gamma_{50}$ or $\Gamma_{05}$, it does not depend on the gauge background $U$, but it connects these two matrices by means of the identities
\bea
\Gamma_{50}(x,y)&=&\sum_z\ep(x,z)\Gamma_{05}(z,y)=\sum_z\Gamma_{05}(x,z)\ep(z,y)
\label{ga50_is_eps_ga05}
\\
\Gamma_{05}(x,y)&=&\sum_z\ep(x,z)\Gamma_{50}(z,y)=\sum_z\Gamma_{50}(x,z)\ep(z,y)
\label{ga05_is_eps_ga50}
\;.
\eea
In passing we note that all occurences of $U_\mu$ in these formulae should be replaced by the smeared gauge field $V_\mu(x)$,
as this greatly reduces the effects of taste symmetry breaking \cite{Blum:1996uf,Orginos:1998ue,Lee:1999zxa,Knechtli:2000ku}.
Last but not least, the staggered action (\ref{def_stag}) is $\ep$-hermitean, i.e.\ $\ep\DS\ep=\DS\dag$ \cite{BOOK_MM}.

\begin{figure}[!tb]
\includegraphics[width=0.48\textwidth]{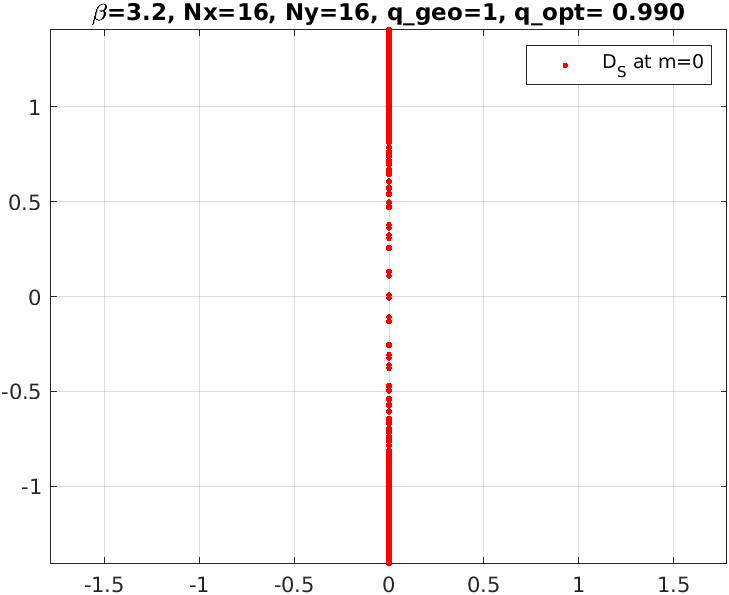}\hfill
\includegraphics[width=0.49\textwidth]{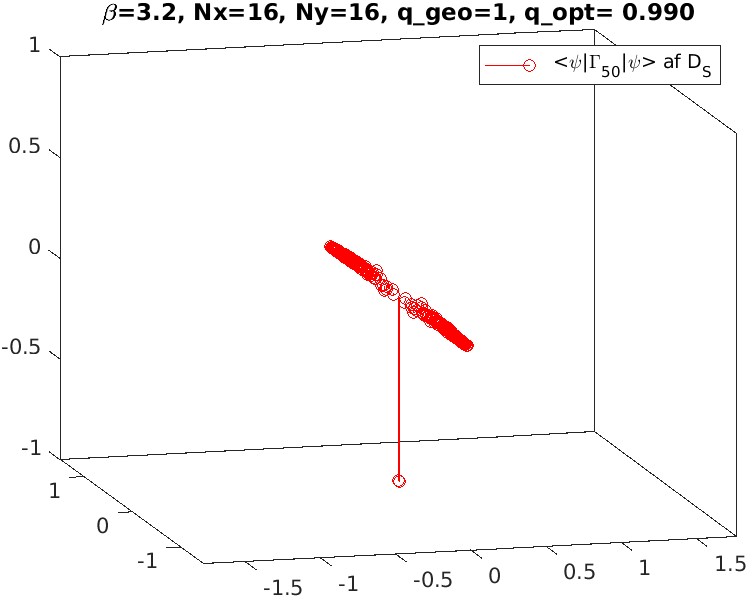}%
\vspace*{-2mm}
\caption{\label{fig:eigstem_stag}\sl
Eigenvalues of the staggered operator on a background with $q=1$ (left),
and ``needle plot'' of the $\Gamma_{50}$-chiralities in the pertinent left-right-eigenvector sandwich (right).
The standard $\ep$-chiralities or $\Gamma_{55}$-chiralities are exactly flat (not shown).}
\vspace*{+2mm}
\includegraphics[width=0.49\textwidth]{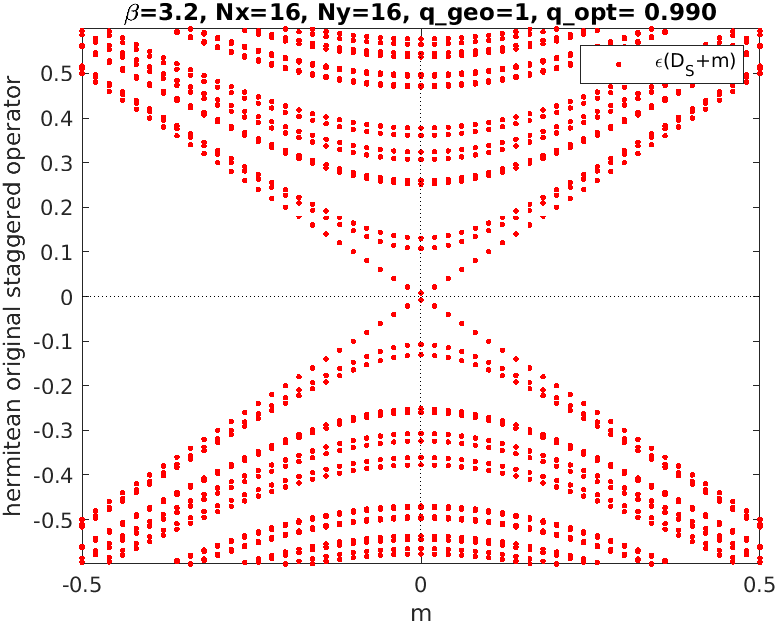}\hfill
\includegraphics[width=0.49\textwidth]{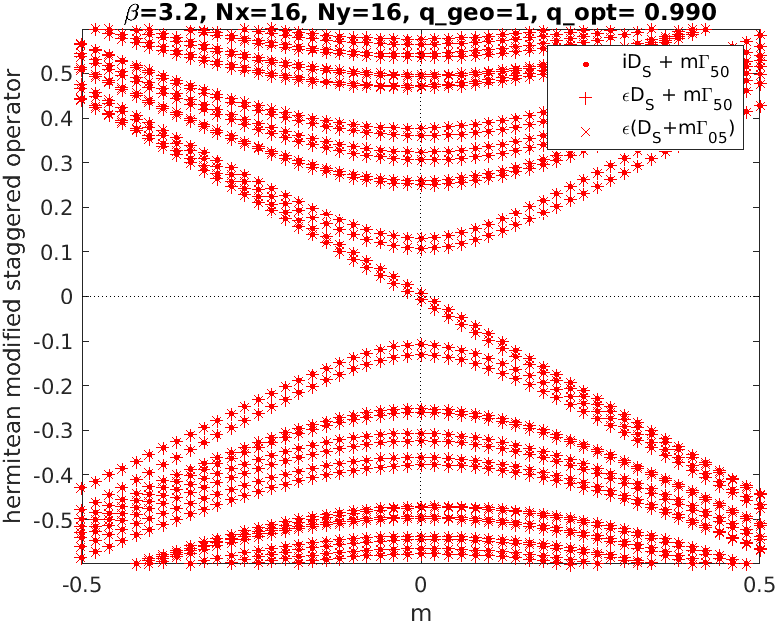}%
\vspace*{-2mm}
\caption{\label{fig:specflow_stag}\sl
Naive spectral flow of the staggered Dirac operator, i.e.\ eigenvalues of $\ep(\DS+m)$ versus $m$ (left),
and eigenvalues of $\ri \DS+m\Gamma_{50}$, $\ep\DS+m\Gamma_{50}$, $\ep(\DS+m\Gamma_{05})$ versus $m$ (right).}
\end{figure}

The eigenvalue spectrum of $\DS$ on the same gauge configuration as in Sec.~\ref{sec:wilsbril} is shown in Fig.~\ref{fig:eigstem_stag}.
The eigenvalues $\la_i$ are purely imaginary, and there is a pairing property $\la\leftrightarrow-\la$ which reflects the $\ep$-hermiticity.
Close inspection reveals that the eigenvalues are two-fold nearly degenerate, i.e.\ each blob in the figure actually represents two nearby eigenvalues.
In particular the blob on the real axis represents the two would-be zero-eigenvalues%
\footnote{\label{foot:stagvalues}
The precise values of the two would-be zero-eigenvalues are $\la\simeq\pm0.00776\ri$.}
expected for topological charge $q=1$ (in general $2|q|$ for $q\in\mathbb{Z}$).
In 4D the near-degeneracy is $4$-fold \cite{Follana:2004sz,Durr:2004as}.

Like in the previous section, we proceed by calculating all eigenvectors of $\DS$.
Since $\DS$ is normal, it suffices to compute the right-eigenvector $|\ps_i\>$ for any $\la_i$, the left-eigenvector is just $\<\ps_i|=(|\ps_i\>)\dag$.
The chirality is defined as expected value of the chirality operator in the sandwich between $\<\ps_i|$ and $|\ps_i\>$.
However, choosing $\ep$ as chirality operator is not a good choice, since $\<\ps_i|\ep|\ps_i\>=0$ holds%
\footnote{This fact occasionally mislead people to believe that ``staggered fermions are blind to topology''.}
for all modes.
Early publications demonstrating (on interacting backgrounds) that one must use $\Gamma_5$ as chirality operator include \cite{Smit:1986fn,Smit:1987jh,Smit:1987fq,Laursen:1990ec,Hands:1990wc}.
Choosing $\Gamma_5$ as defined in (\ref{def_gamma5}) and representing the chirality $\<\ps_i|\Gamma_5|\ps_i\>$ as a needle over $\la_i\in\mathbb{C}$
yields the right panel in Fig.~\ref{fig:eigstem_stag}.
The two would-be zero-modes have a chirality close to $-1$, all other modes have chiralities close to zero.
Last but not least, we verified that $\Xi_5$ is insensitive to topology, i.e.\ $\<\ps_i|\Xi_5|\ps_i\>\simeq0$ for all modes.
This difference is crucial; it underpins the workings of the staggered flavor interpretation
(see Refs.~\cite{Sharatchandra:1981si,KlubergStern:1983dg,Golterman:1984cy,Smit:1986fn,Blum:1996uf,Orginos:1998ue,Lee:1999zxa,Knechtli:2000ku} and Ref.~\cite{BOOK_MM} for a guide to the literature).

The spectral flow of the staggered operator is shown in Fig.~\ref{fig:specflow_stag}.
A naive analog to $\gaf(\DW+m)$ would be $\ep(\DS+m)$, but this choice yields no eigenvalue crossing.
The more faithful analog is $\HS=\ep\DS+m\Gamma_{50}$, since $\Gamma_{50}\DS$ is not hermitean.
Since $\mr{spec}(\ep\DS)=\mr{spec}(\ri\DS)$, also $\ri\DS+m\Gamma_{50}$ works fine.
Incidentally, this was the first proposal by Adams to generate a staggered spectral flow \cite{Adams:2009eb}, and it was also used in Ref.~\cite{Azcoiti:2014pfa}.
Finally, due to the property (\ref{ga50_is_eps_ga05}, \ref{ga05_is_eps_ga50}), the faithful choice is identical to $\HS=\ep(\DS+m\Gamma_{05})$ which holds a preview of the Adams operator (see below).
These three (non-naive) choices are seen to yield identical results (with two down-crossings, as expected for a two-species operator and $q=+1$).

\begin{figure}[!tb]
\includegraphics[width=0.48\textwidth]{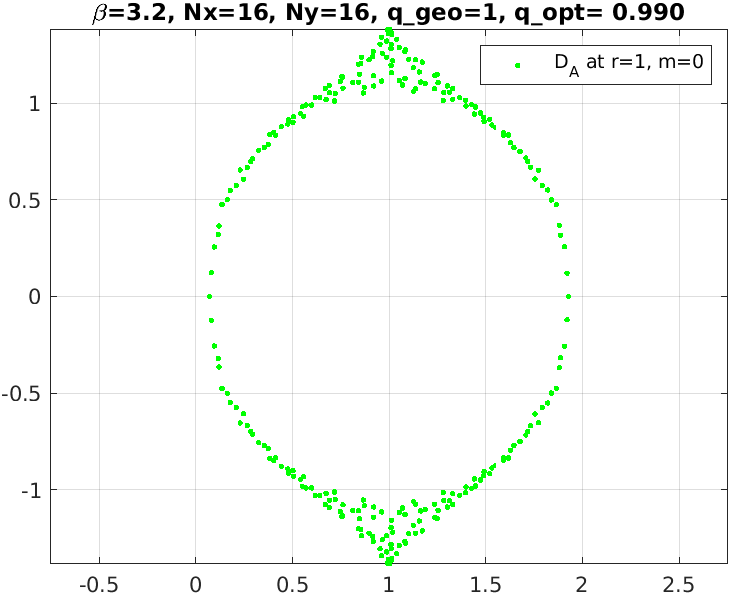}\hfill
\includegraphics[width=0.49\textwidth]{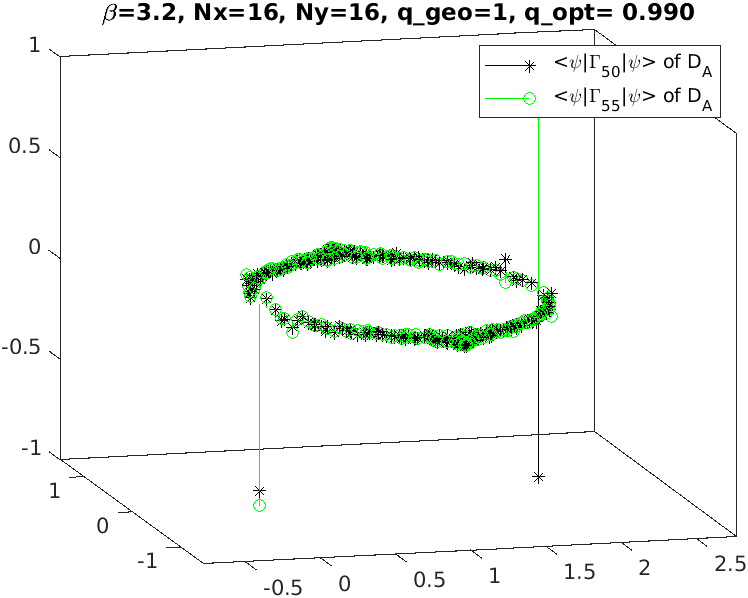}%
\vspace*{-2mm}
\caption{\label{fig:eigstem_adam}\sl
Eigenvalues of the Adams operator on a background with $q=1$ (left),
and ``needle plots'' of the $\Gamma_{50}$-chiralities (black stars) and $\Gamma_{55}$-chiralities
(open green circles) in the pertinent left-right-eigenvector sandwich (right).}
\vspace*{+2mm}
\includegraphics[width=0.49\textwidth]{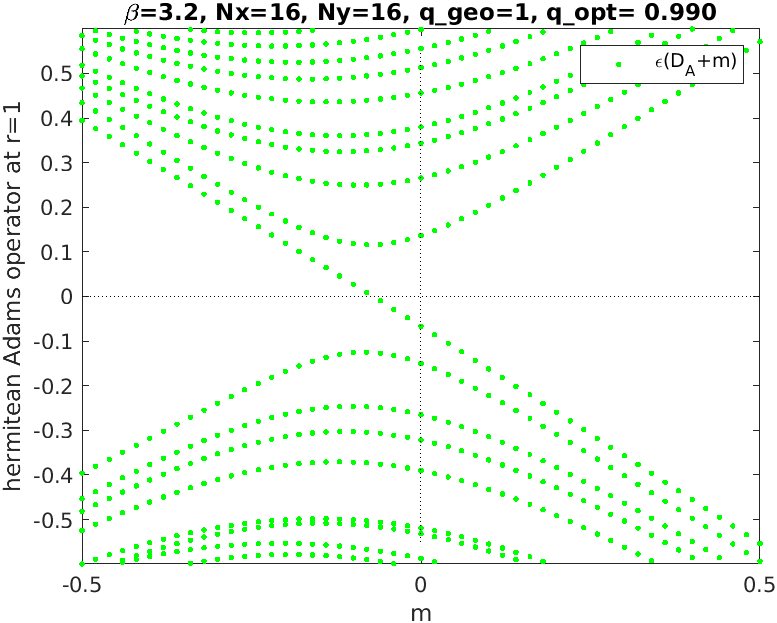}\hfill
\includegraphics[width=0.49\textwidth]{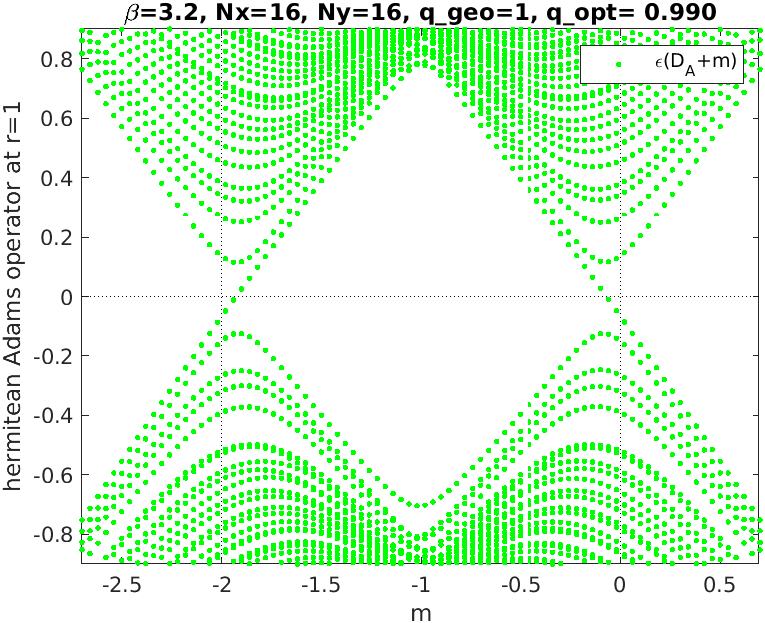}%
\vspace*{-2mm}
\caption{\label{fig:specflow_adam}\sl
Spectral flow of the Adams operator, i.e.\ eigenvalues of $\ep(\DA+m)$ versus $m$.
Relevant part near $m=0$ (left) and panoramic view (right).}
\end{figure}

Armed with this insight, we are in a position to esteem the ingenuity of the operator
\beq
\DA(x,y)=\DS(x,y)+\frac{r}{a}(1\pm\Gamma_{05})
\label{def_adam}
\eeq
proposed by Adams \cite{Adams:2010gx}.
The choice of sign inside the parentheses depends on the conventions underpinning (\ref{def_Gamma5}, \ref{def_Xi5}).
In practice one will choose it such that the ``needle'' in the physical branch (see below) points in the same direction
as in Figs.~\ref{fig:eigstem_wils}, \ref{fig:eigstem_bril} and \ref{fig:eigstem_stag}.
In other words, the Adams proposal is to use $\Gamma_{05}=\Xi_5$ to induce a separation between the physical mode%
\footnote{In 4D the Adams term $\frac{r}{a}(1\pm\Gamma_{05})$ causes a splitting between two near-degenerate physical modes and two near-degenerate doublers,
see \cite{Hoelbling:2010jw,deForcrand:2012bm,Durr:2013gp} for illustrations in the free-field case and the interacting case.}
and the doubler mode.
Like with the other operators, we introduce a Wilson-like deformation parameter $r$ in (\ref{def_adam}).
The Adams operator (\ref{def_adam}) is $\ep$-hermitean, i.e.\ $\ep\DA\ep=\DA\dag$.

The eigenvalues of $\DA$ are shown in Fig~\ref{fig:eigstem_adam} (still using the same $q=1$ background as before).
The eigenvalue spectrum is symmetric about the real axis (owing to the $\ep$-hermiticity), but there is no reflection symmetry about $\mr{Re}(z)=1$.
Like in the Wilson/Brillouin case there is a single (exactly real) would-be zero-mode in the physical branch (and another one near $2r$).
We compute the left-eigenvector $\<\ps_i|$ and the right-eigenvector $|\ps_i\>$ for each eigenvalue%
\footnote{In doing so we keep in mind that $\DA$ is not normal, $[\DA,\DA\dag]\neq0$, see footnote~\ref{foot:normality}) for details.}
$\la_i$ and plot the chirality $\<\ps_i|\ep|\ps_i\>$ as a needle (with green circle) at position $\la_i\in\mathbb{C}$.
It is nearly $-1$ at $\la=0.0706$ (in the physical branch) and nearly $+1$ at $\la=1.9276$ (in the doubler branch).
Why is $\ep$ the correct chirality operator for $\DA$ ?
The answer was given in the discussion of the (working) spectral flow plot for $\DS$ in Fig.~\ref{fig:specflow_stag}.
The pluses and crosses give the eigenvalues of
\beq
\ep\DS+m\Gamma_{50}=\ep(\DS+m\Gamma_{05})
\eeq
where the equality follows from (\ref{ga50_is_eps_ga05}, \ref{ga05_is_eps_ga50}).
This way Adams managed to have the operator $\ep$, which induces the hermiticity property of $\DS$, in front,
and one recognizes that the term in parentheses is just a shifted version of $\DA$ (with $m$ taking the role of $r$).
Our figure also illustrates what happens if one measures the chirality of $\DA$ with the wrong chirality operator $\Gamma_5$ (needles with black stars).
This time either branch has a downward-pointing needle, and upon letting $r\to0$ the situation smoothly turns into the staggered ``needle plot'' shown in Fig.~\ref{fig:eigstem_stag}.
Conversely, the Adams choice of chirality, $\ep$, is not an option in the staggered case, since upon letting $r\to0$ the two oppositely oriented needles (green circles)
would annihilate in this limit and yield an entirely flat chirality plot (as discussed in the staggered paragraph above).

The spectral flow plot for the Adams operator, i.e.\ the eigenspectrum of $\ep(\DA+m)$ versus $m$, is shown in Fig.~\ref{fig:specflow_adam}.
The left panel shows a single down-crossing (as expected for an undoubled operator and $q=1$), the right one clarifies that there is no net crossing.
Note that the left panel of Fig.~\ref{fig:specflow_adam} is not identical to the right panel of Fig.~\ref{fig:specflow_stag} (in the former case one sees the effect
of additive mass renormalization, like for $\DW$ or $\DB$, while in the latter case the crossing of the two physical modes is symmetric about $m=0$).
The relationship between these two plots is more subtle -- the parameter $m$ in Fig.~\ref{fig:specflow_stag} is a disguised version of $r$ in the Adams operator (\ref{def_adam}) at $am=-r$,
while in Fig.~\ref{fig:specflow_adam} we have $r=1$ fixed, and the variable $m$ is really a mass.


\section{Naive fermions without and with species-lifting term\label{sec:naiv}}


The naive Dirac operator at zero bare mass is defined as
\beq
\DN(x,y)=\sum_\mu \ga_\mu \nab_\mu(x,y)
\label{def_naiv}
\eeq
where the anti-hermitean behavior $\nab_\mu\dag=-\nab_\mu$ makes the operator $\gaf$-hermitean, i.e.\ $\gaf \DN \gaf=\DN\dag$.
In the free-field limit this operator assumes a diagonal form in momentum space,
\beq
\DN(p)=\ri \sum_\mu \ga_\mu \frac{1}{a}\sin(ap_\mu)
=\ri\sum_\mu \ga_\mu \bar{p}_\mu
\quad \mbox{with} \quad \bar{p}_\mu=\frac{1}{a}\sin(ap_\mu)
\label{momrep_naiv}
\eeq
which again highlights the anti-hermitean nature of the derivative (momentum) term.

\begin{figure}[!tb]
\includegraphics[width=0.48\textwidth]{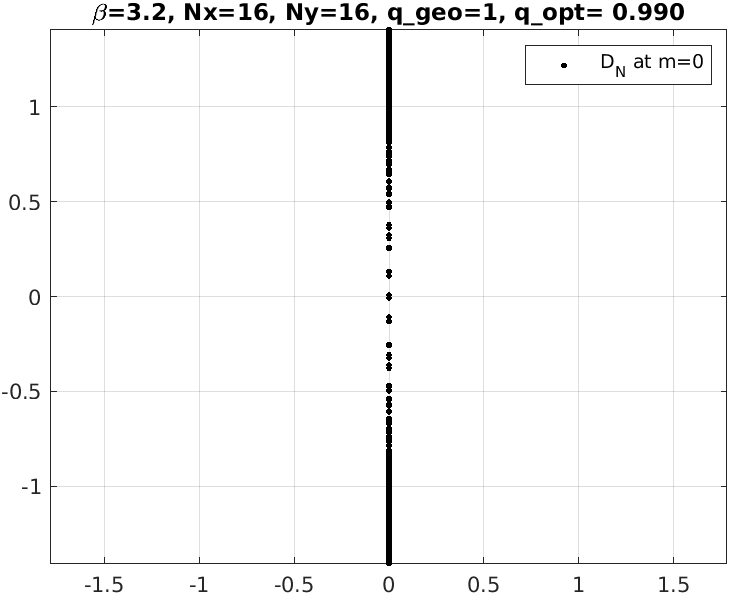}\hfill
\includegraphics[width=0.49\textwidth]{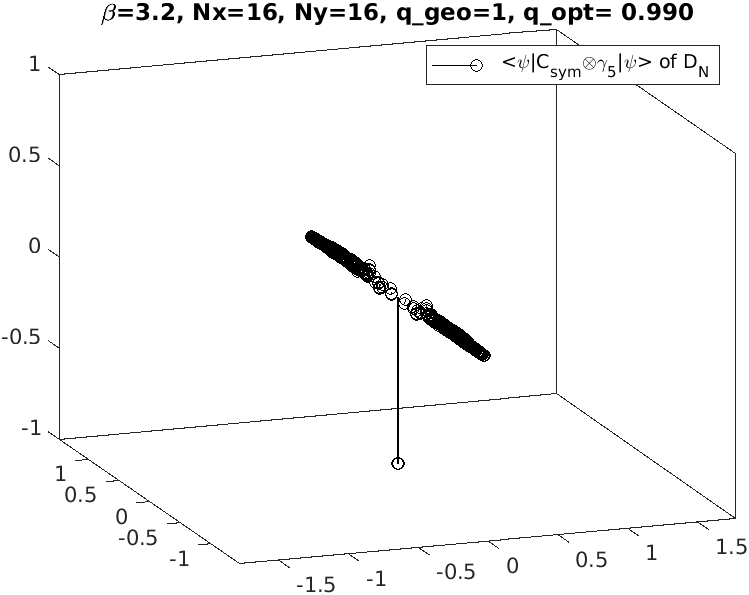}%
\vspace*{-2mm}
\caption{\label{fig:eigstem_naiv}\sl
Eigenvalues of the naive Dirac operator on a background with $q=1$ (left),
and ``needle plot'' of the $C_\mr{sym}\otimes\gaf$-chiralities in the pertinent left-right-eigenvector sandwich (right).
The $[\frac{1}{2}(C_1+C_2)^2-1]\otimes\gaf$-chiralities look similar, the $\gaf$-chiralities are zero (not shown).}
\vspace*{+2mm}
\includegraphics[width=0.49\textwidth]{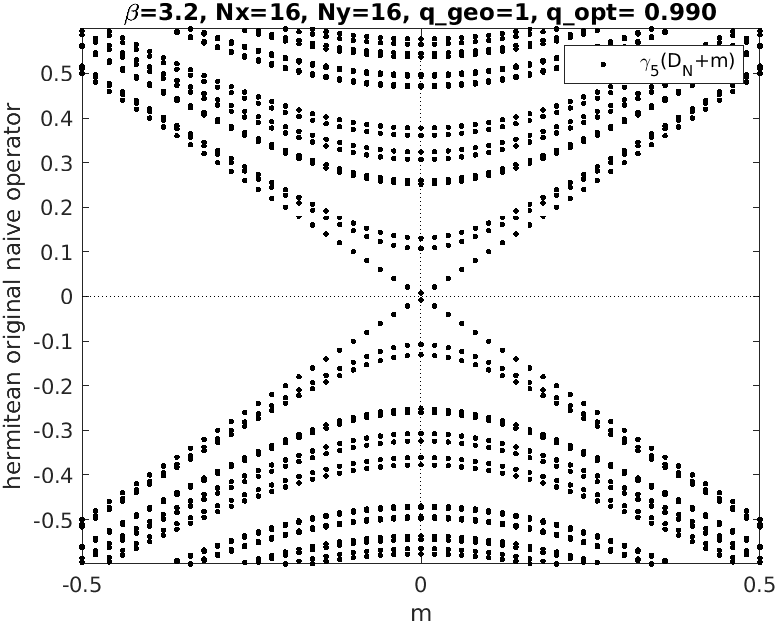}\hfill
\includegraphics[width=0.49\textwidth]{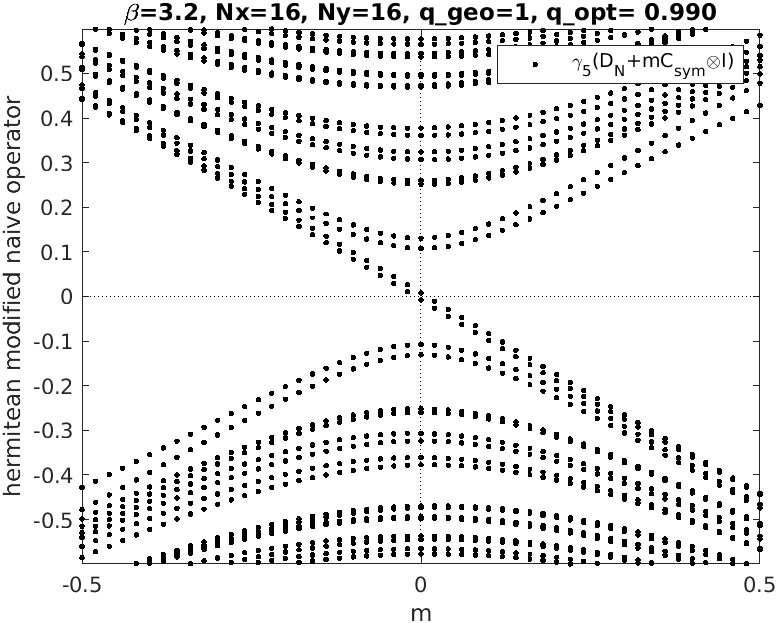}%
\vspace*{-2mm}
\caption{\label{fig:specflow_naiv}\sl
Spectral flow of the naive Dirac operator, i.e.\ eigenvalues of $\gaf(\DN+m)$ versus $m$ (left),
and eigenvalues of $\gaf(\DN+mC_\mr{sym}{\otimes}1)$ versus $m$ (right); every point two-fold degerate.}
\end{figure}

Based on the 1-hop operators $C_\mu(x,y)$ given in App.~\ref{app:notation} one defines the $d$-hop operator
\beq
C_\mr{sym}=\frac{1}{d!}\sum_\mr{perm} C_1 C_2 ... C_d
\eeq
or specifically $C_\mr{sym}=\frac{1}{2}\{C_1,C_2\}$ in $d=2$ dimensions, and $C_\mr{sym}=\frac{1}{24}[C_1C_2C_3C_4+\mr{perm}]$ in $d=4$ dimensions.
Furthermore, there is the operator $\frac{1}{2}(C_1+C_2)^2=C_\mr{sym}+\frac{1}{2}(C_1^2+C_2^2)$ in 2D.
Note that these operators depend on the gauge background (just as $\Gamma_5$ and $\Xi_5$ in Sec.~\ref{sec:stagadam} did).

The eigenvalue spectrum of the naive Dirac operator $\DN$ on our $q=1$ gauge background is shown in Fig.~\ref{fig:eigstem_naiv}.
The dots coincide%
\footnote{The blob on the real axis is at $\la\simeq\pm0.00776\ri$, cf.\ footnote~\ref{foot:stagvalues}, and each eigenvalue is two-fold degenerate.}
with the staggered dots in Fig.~\ref{fig:eigstem_stag}, but there is an additional (exact) two-fold%
\footnote{In 4D the exact degeneracy is four-fold, but after this degeneracy has been removed the eigenvalue spectrum would
again coincide with the staggered eigenvalue spectrum (which in 4D has a four-fold near-degeneracy).}
degeneracy.
The naive operator is ``blind to topology'' if one uses $\gaf$ to define chirality, since $\<\ps_i|\gaf|\ps_i\>=0$ holds for each eigenmode $\ps_i$ of $\DN$.
This situation is reminiscent of choosing $\ep$ as chirality operator in the staggered case; this gave $\<\ps_i|\ep|\ps_i\>=0$ for each eigenmode $\ps_i$ of $\DS$.
However, in the staggered case the situation changed by switching to $\Gamma_5$ as the chirality operator,
and one wonders whether using $C_\mr{sym}\otimes\gaf$ or $\frac{1}{2}(C_1+C_2)^2\otimes\gaf$ might bring a similar change for the naive Dirac operator.
The right panel displays the chirality $\<\ps_i|C_\mr{sym}\otimes\gaf|\ps_i\>$ as a needle at position $\la_i\in\mathbb{C}$ for each $i$.
Hence $C_\mr{sym}\otimes\gaf$ works perfectly as chirality operator; we find four modes%
\footnote{In 4D there are $16$ needles on a background with $q=\pm1$; in general $2^d$ in $d$ space-time dimensions.}
reaching almost down to $-1$, as expected for a fermion operator which encodes for four continuum species.
We also tried $\frac{1}{2}(C_1+C_2)^2\otimes\gaf$, and the respective ``needle plot'' is hard to distinguish from the one in Fig.~\ref{fig:eigstem_naiv}.
Still, there is a subtle difference%
\footnote{In the subspace of would-be zero-modes the chiralities $C_\mr{sym}{\otimes}\gaf$ and $\frac{1}{2}(C_1+C_2)^2{\otimes}\gaf$ take the form
\bdm
C_\mr{sym}{\otimes}\gaf \doteq
\begin{pmatrix}
   -0.9148 \!&\!   0.0    \!&\!  -0.0013 \!&\!   0.0    \\
    0.0    \!&\!  -0.9148 \!&\!   0.0    \!&\!  -0.0013 \\
   -0.0013 \!&\!   0.0    \!&\!  -0.9148 \!&\!   0.0    \\
    0.0    \!&\!  -0.0013 \!&\!   0.0    \!&\!  -0.9148 \\
\end{pmatrix}
\!,\;
\frac{1}{2}(C_1+C_2)^2{\otimes}\gaf \doteq
\begin{pmatrix}
     -0.915 \!&\!     0.0   \!&\!    -0.939 \!&\!     0.0   \\
      0.0   \!&\!    -0.915 \!&\!     0.0   \!&\!     0.936 \\
     -0.939 \!&\!     0.0   \!&\!    -0.915 \!&\!     0.0   \\
      0.0   \!&\!     0.936 \!&\!     0.0   \!&\!    -0.915 \\
\end{pmatrix}
\edm
and considering elements $O(10^{-3})$ as zero, the former matrix is diagonal, while the latter one is not.
The attentive reader may think of definining new basis vectors, e.g.\ ``(first+third)'' or ``(second+fourth)'' in this subspace, with normalization $1/\sqrt{2}$.
The first matrix would be unchanged, while the second one would become close to diagonal.
However, this proposal ignores that the first two eigenvectors belong to $\la\simeq+0.00776\ri$, and the latter two to $\la\simeq-0.00776\ri$.
Hence, after the proposed rotation, this is no longer an eigenbasis of $\DN$.}
in the sense that the former operator is almost diagonal on the subspace spanned by the would-be zero-modes, while the latter one is not.
But the chirality operator $[\frac{1}{2}(C_1+C_2)^2-1]\otimes\gaf$ ameliorates the situation again.
Its ``needle plot'' still looks like in Fig.~\ref{fig:eigstem_naiv}, while it is again close to diagonal on the subspace spanned by the would-be zero-modes.

The spectral flow plots for the naive action are presented in Fig.~\ref{fig:specflow_naiv}.
Using $\gaf$ as chirality operator, $\DN$ shows no crossing; the eigenvalues of $\gaf(\DN+m)$ are symmetric under $m\leftrightarrow-m$.
This situation is analogous to the staggered case with $\ep$ as chirality operator.
Choosing instead $C_\mr{sym}\otimes\gaf$ as chirality operator, the situation changes.
In view of the exact two-fold degeneracy%
\footnote{Choosing instead $[\frac{1}{2}(C_1+C_2)^2-1]\otimes\gaf$ as chirality operator, one finds a similar crossing picture, but
the eigenvalues of $\gaf(\DN+m[\frac{1}{2}(C_1+C_2)^2-1]\otimes1)$ are two-fold near-degenerate rather than exactly degenerate.}
in the eigenvalues of $\gaf(\DN+mC_\mr{sym}\otimes1)$, there are (in total) four down-crossings, as expected for a four-species formulation (in 2D) and $q=1$.
Modulo this degeneracy, the right panel bears strong similarity with the respective staggered panel, i.e.\ eigenvalues of $\ep(\DS+m\Gamma_{50})$.

\begin{figure}[!tb]
\includegraphics[width=0.48\textwidth]{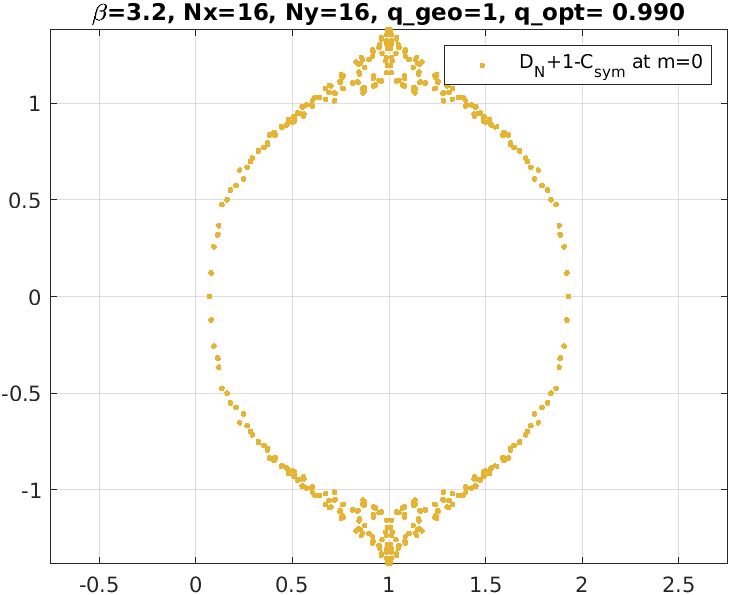}\hfill
\includegraphics[width=0.49\textwidth]{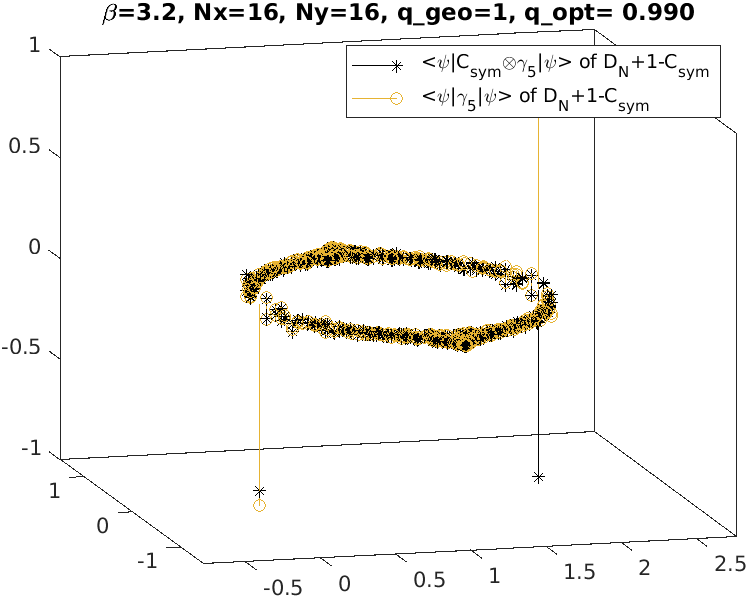}%
\vspace*{-2mm}
\caption{\label{fig:eigstem_like}\sl
Eigenvalues of the ``Adams-like'' Dirac operator $\DN+1-C_\mr{sym}$ on a background with $q=1$ (left),
and ``needle plots'' of the $C_\mr{sym}\otimes\gaf$-chiralities (black stars) and $\gaf$-chiralities (golden circles) in the pertinent left-right-eigenvector sandwich (right).}
\vspace*{+2mm}
\includegraphics[width=0.49\textwidth]{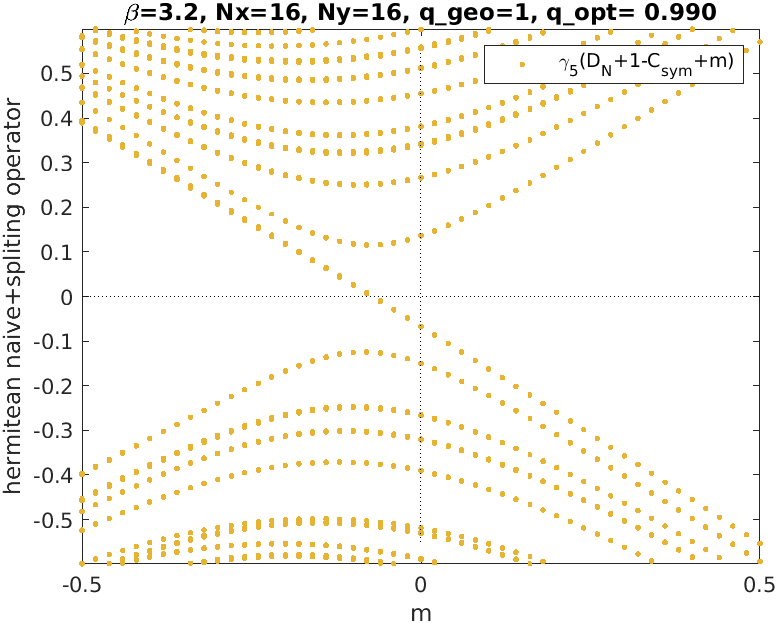}\hfill
\includegraphics[width=0.49\textwidth]{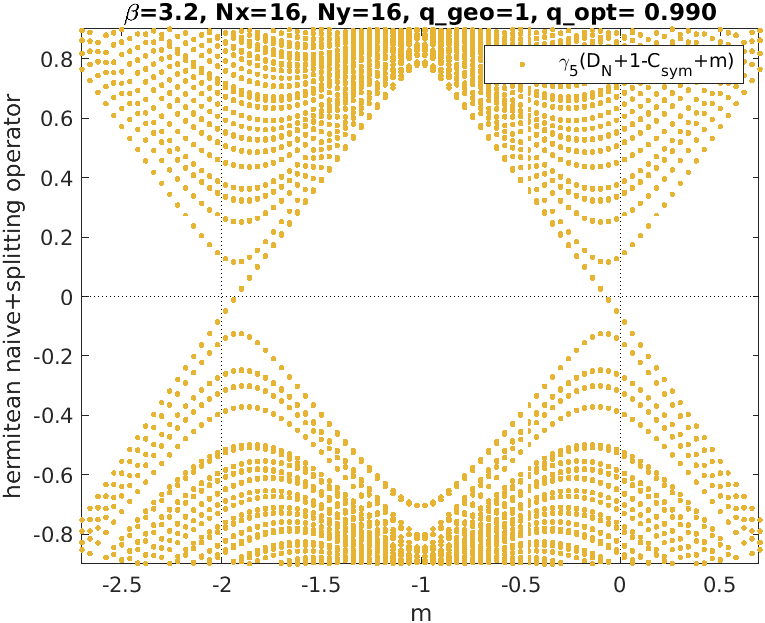}%
\vspace*{-2mm}
\caption{\label{fig:specflow_like}\sl
Spectral flow of the ``Adams-like'' Dirac operator, i.e.\ eigenvalues of $\gaf(\DN+1-C_\mr{sym}+m)$ versus $m$.
Relevant part near $m=0$ (left) and panoramic view (right).}
\end{figure}

Given the similarity between the right panels of Fig.~\ref{fig:specflow_stag} and Fig.~\ref{fig:specflow_naiv}, 
and bearing in mind the process which led to the construction of $\DA$ in (\ref{def_adam}), one defines the ``Adams-like'' operator
\beq
D_\mr{like}(x,y)=\sum_\mu \ga_\mu \nab_\mu(x,y) + \frac{r}{a}(1\pm C_\mr{sym})_{x,y}
\label{def_like}
\eeq
which realizes a ``2+2'' taste-splitting in $d=2$ dimensions, and a ``8+8'' splitting in $d=4$ dimensions \cite{Creutz:2010bm}.
In either dimension the splitting is \emph{consistent with chirality}, i.e.\ the physical modes share one chirality and all doubler modes have opposite chirality.
This feature holds true for Adams fermions, provided their chirality is measured with $\Gamma_{55}=\ep$.
For the operator (\ref{def_like}) it holds true in conjunction with the standard $\gaf$-definition of the chirality.

The eigenvalues $\la_i\in\mathbb{C}$ of the ``Adams-like'' operator (\ref{def_like}) are shown in Fig.~\ref{fig:eigstem_like}.
The term $1-C_\mr{sym}$ separates the two branches nicely, and all eigenvalues are two-fold near-degenerate.
After removing this near-degeneracy, the eigenvalue spectrum bears a striking similarity with the one of the Adams operator,
see Fig.~\ref{fig:eigstem_adam}, with a comparable size of the additive mass shift.
The would-be zero-modes are exactly real, and the eigenvalues associated with non-chiral modes come in complex conjugate pairs, owing to the $\gaf$-hermiticity of $D_\mr{like}$.
Unlike in Fig.~\ref{fig:eigstem_adam}, there is a reflection symmetry about $\mr{Re}(z)=1$.
We compute the right-eigenvector $|\ps_i\>$ and the left-eigenvector $\<\ps_i|$ for each eigenvalue and plot the
(correct) chirality $\<\ps_i|\gaf|\ps_i\>$ as a needle (golden circle) at position $\la_i\in\mathbb{C}$.
Two modes in the physical branch reach almost down to $-1$, and two modes in the doubler branch reach almost up to $+1$.
Our figure also illustrates the result of combining the action (\ref{def_like}) with the wrong chirality operator $C_\mr{sym}\otimes\gaf$ (needles with black stars).
In this case either branch has downward-pointing needles, and upon letting $r\to0$ the situation smoothly turns into the naive chirality plot shown in Fig.~\ref{fig:eigstem_naiv}.
Conversely, the correct choice of chirality, $\gaf$, is not an option in the naive case, since upon letting $r\to0$ the two oppositely oriented needles (golden circles)
would annihilate in this limit and yield an entirely flat chirality plot (as discussed in the naive paragraph above).

The spectral flow plot of the operator (\ref{def_like}), i.e.\ the eigenvalues of $\gaf(\DN+1-C_\mr{sym}+m)$ versus $m$, is shown in Fig.~\ref{fig:specflow_like}.
Each dot represents two nearly degenerate eigenvalues.
Considering the situation in the vicinity of $m=0$ one finds two down-crossings, as expected for a two-species formulation and $q=1$.
The right panel shows that there is no net crossing, and the situation is symmetric about $m=-1$ (which holds only approximately in the Adams case).


\section{Central-branch fermions and descendants\label{sec:cebr}}


The ``central-branch'' Dirac operator at vanishing bare mass is defined as \cite{Chowdhury:2013ux,Misumi:2019jrt,Misumi:2020eyx}
\beq
D_\mr{cb}(x,y)=\sum_\mu \ga_\mu \nab_\mu(x,y)
+\frac{r}{a}\Big[-\frac{a^2}{2} \sum_\mu \lap_\mu-dI\Big]_{x,y}
\label{def_cb}
\eeq
where $I$ denotes the identity in position space.
In the notation of App.~\ref{app:notation} the square bracket is $-\sum_\mu C_\mu$, see (\ref{def_cmu}),
and it is evaluated at $(x,y)$ whereupon $I(x,y)=\de_{x,y}$.
In the free-field limit the ``central branch'' operator assumes a diagonal form in momentum space
\bea
D_\mr{cb}(p)&=&\ri \sum_\mu \ga_\mu \frac{1}{a}\sin(ap_\mu)+\frac{r}{a}\sum_\mu \{0-\cos(ap_\mu)\}
\nonumber
\\
&=&\ri \sum_\mu \ga_\mu \bar{p}_\mu
+\frac{r}{a}\Big[\frac{a^2}{2} \sum_\mu \hat{p}_\mu^2-d\Big]
\label{momrep_cb}
\eea
which confirms that it is a shifted version of the Wilson operator (\ref{def_wils}, \ref{momrep_wils}).

We refrain from showing a plot of the eigenvalues of $D_\mr{cb}$, since it is just a copy of Fig.~\ref{fig:eigstem_wils}, but shifted by $2$ units to the left.
This formulation leads to $2$ species in 2D or $6$ in 4D.
If chirality is measured by the usual $\gaf$ operator, the physical species \emph{share one chirality}.
There is no additive mass renormalization, but one should not be fooled to believe%
\footnote{\label{foot:taste}
It pays to consider the symmetries of the underlying taste structure \cite{Yumoto:2021fkm}.
The Wilson lifting term is $\DW-\DN=-\frac{a}{2}\sum\lap_\mu=\frac{1}{a}(-\sum C_\mu+dI)$, where $I$ is the identity.
Since $\gaf$ commutes with both $C_\mu$ and $I$, we have $\{\gaf,-\sum C_\mu\}=-2\gaf\sum C_\mu$ and $\{\gaf,I\}=2\gaf$, so both $\sum C_\mu$ and $I$ break chiral symmetry.
Still, there is an important difference between these operators.
Let $\tau_{\mu,x}=(-1)^{x_\mu}\ri\gamma_\mu\gaf$ be the generator of the (removable) taste symmetry of naive fermions,
and consider a taste rotation $\psi_x \to \tau_{\mu,x} \psi_x$ and $\bar\psi_x \to \bar\psi_x\tau_{\mu,x}$.
Then $\DN(x,y)=\tau_{\mu,x}\DN(x,y)\tau_{\mu,y}$ and $I(x,y)=\tau_{\mu,x}I(x,y)\tau_{\mu,y}$ transform in the same way.
But the hopping part of the Wilson term transforms as
$C_\mu(x,y)=-\tau_{\mu,x}C_\mu(x,y)\tau_{\mu,y}$ and
$C_\nu(x,y)=+\tau_{\mu,x}C_\nu(x,y)\tau_{\mu,y}$ for $\mu\neq\nu$.
Hence, $I$ and $\sum C_\mu(x,y)$ in $\DW$ and $D_\mr{cb}$ do not share the full set of symmetries, and renormalize differently.
In summary, the central branch term $-\sum C_\mu$ does not mix with the identity, but still breaks chiral symmetry.}
that there is true chiral symmetry.
In fact, a close look at Fig.~\ref{fig:eigstem_wils} reveals that the eigenvalues in the central branch are slightly ``fuzzed'' in the horizontal direction,
in contradistinction to actions with a remnant chiral symmetry like $\DS$, $\DN$, $\DKW$ and $\DBC$ (cf.\ Secs.~\ref{sec:stagadam}, \ref{sec:naiv}, \ref{sec:kawi}, \ref{sec:bocr}).

\begin{figure}[!tb]
\includegraphics[width=0.48\textwidth]{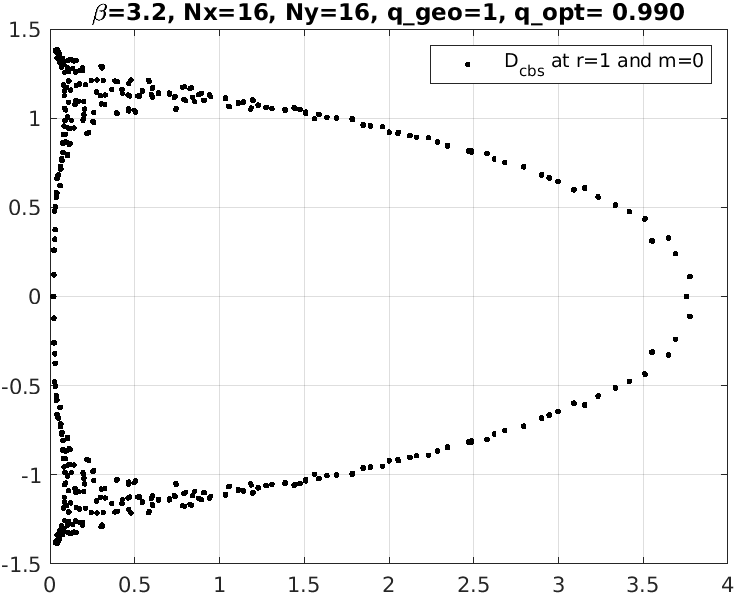}\hfill
\includegraphics[width=0.49\textwidth]{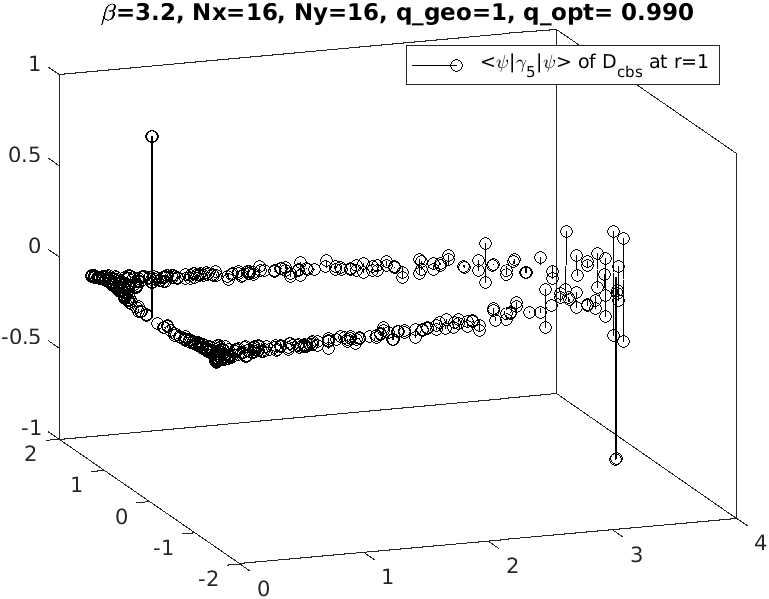}%
\caption{\label{fig:eigstem_cbsq}\sl
Eigenvalues of the ``central-branch-squared'' Dirac operator at $r=1$ on a background with $q=1$ (left),
and ``needle plot'' of the $\gaf$-chiralities in the pertinent left-right-eigenvector sandwich (right).}
\vspace*{+2mm}
\includegraphics[width=0.49\textwidth]{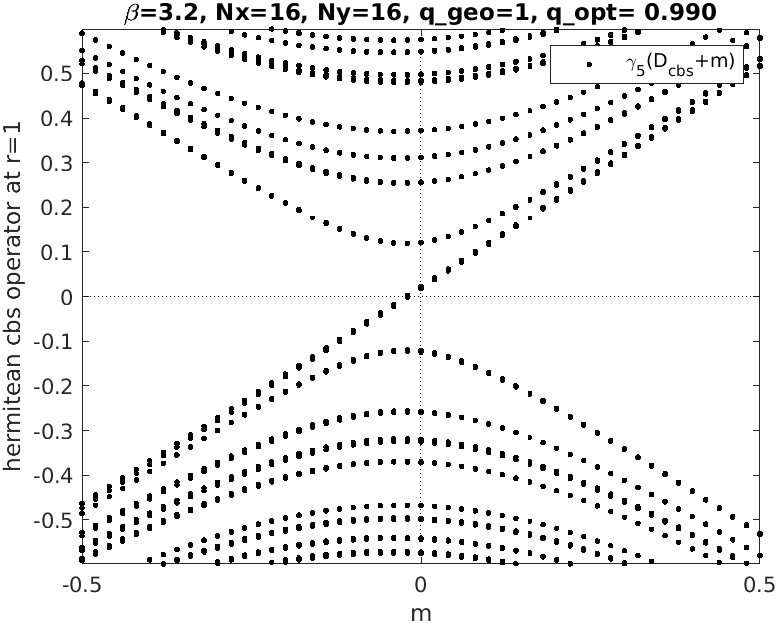}\hfill
\includegraphics[width=0.49\textwidth]{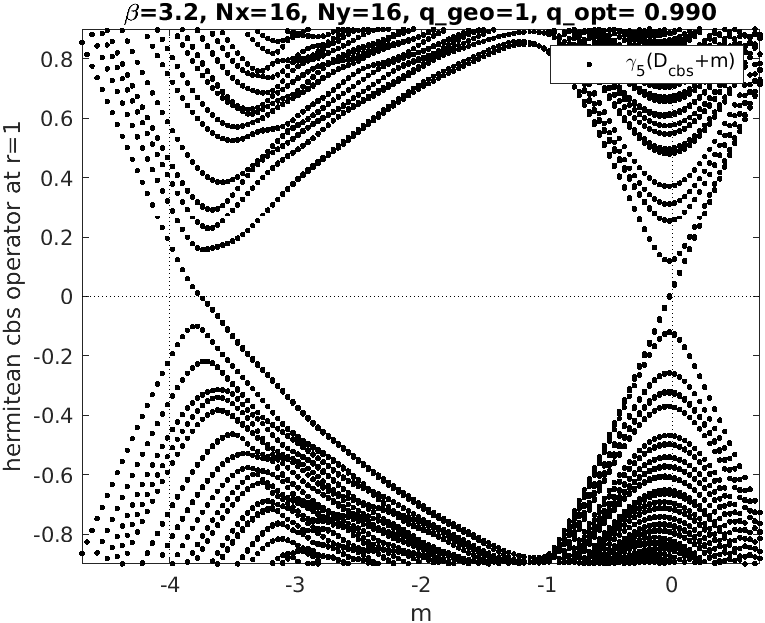}%
\vspace*{-2mm}
\caption{\label{fig:specflow_cbsq}\sl
Spectral flow of the ``central-branch-squared'' Dirac operator at $r=1$, i.e.\ eigenvalues of $\gaf(D_\mr{cbs}+m)$ versus $m$.
Relevant part near $m=0$ (left) and panoramic view (right).}
\end{figure}

The ``central-branch-squared'' Dirac operator (briefly mentioned in Ref.~\cite{Durr:2021gma}) is defined as
\beq
D_\mr{cbs}(x,y)=\sum_\mu \ga_\mu \nab_\mu(x,y)
+\frac{r}{a}\Big[-\frac{a^2}{2} \sum_\mu \lap_\mu-dI\Big]_{x,y}^2
\label{def_cbs}
\eeq
and in the free-field limit it assumes a diagonal form in momentum space
\bea
D_\mr{cbs}(p)&=&\ri \sum_\mu \ga_\mu \frac{1}{a}\sin(ap_\mu)+\frac{r}{a}\Big[\sum_\mu \{0-\cos(ap_\mu)\}\Big]^2
\nonumber
\\
&=&\ri \sum_\mu \ga_\mu \bar{p}_\mu
+\frac{r}{a}\Big[\frac{a^2}{2} \sum_\mu \hat{p}_\mu^2-d\Big]^2
\label{momrep_cbs}
\eea
which indicates that indeed only the lifting term in (\ref{def_cb}, \ref{momrep_cb}) is squared.
In 2D this operator has two branches with ``2+2'' multiplicities, in 4D it has three branches with ``6+8+2'' multiplicities.

\begin{figure}[!tb]
\includegraphics[width=0.48\textwidth]{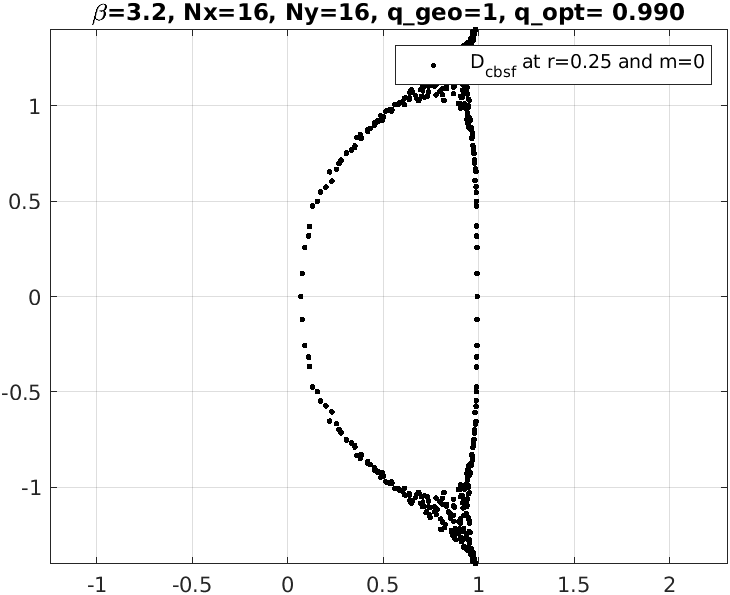}\hfill
\includegraphics[width=0.49\textwidth]{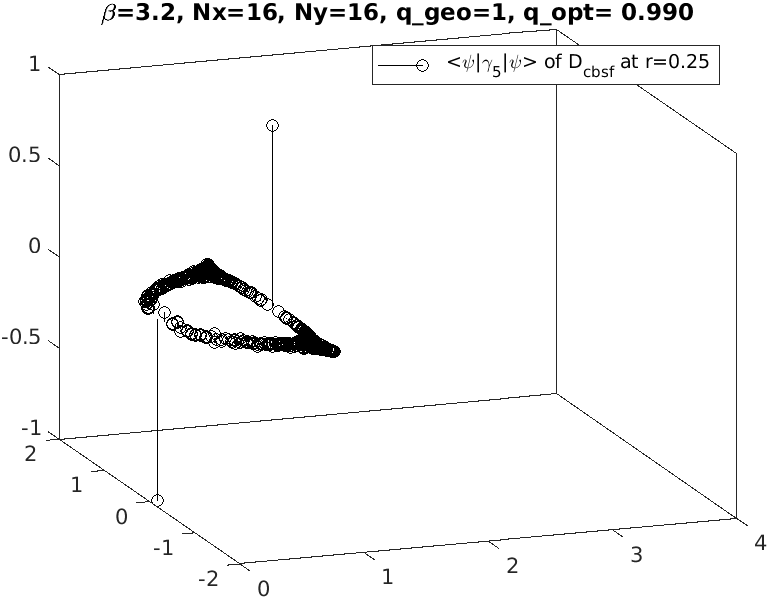}%
\vspace*{-2mm}
\caption{\label{fig:eigstem_cbsf}\sl
Eigenvalues of the ``central-branch-squared-and-flipped'' Dirac operator at $r=\frac{1}{4}$ on a background with $q=1$ (left),
and ``needle plot'' of the $\gaf$-chiralities in the pertinent left-right-eigenvector sandwich (right).}
\vspace*{+2mm}
\includegraphics[width=0.49\textwidth]{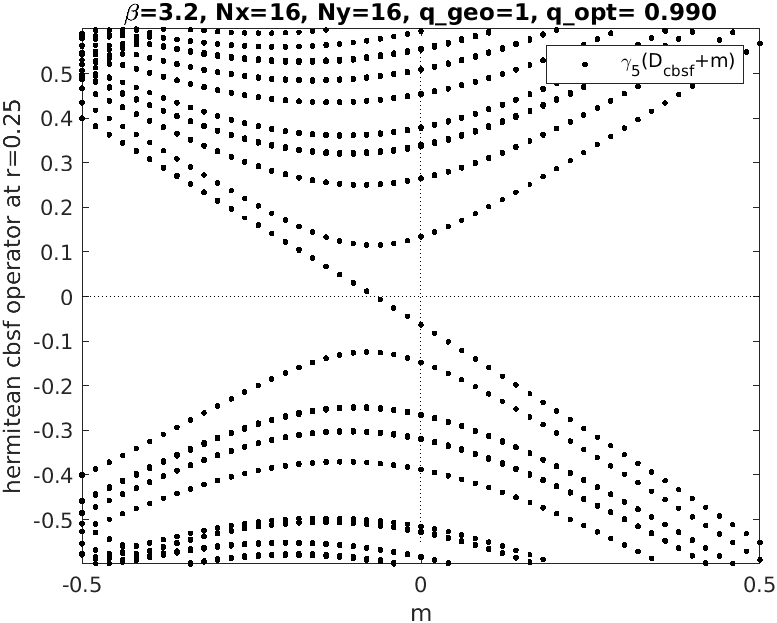}\hfill
\includegraphics[width=0.49\textwidth]{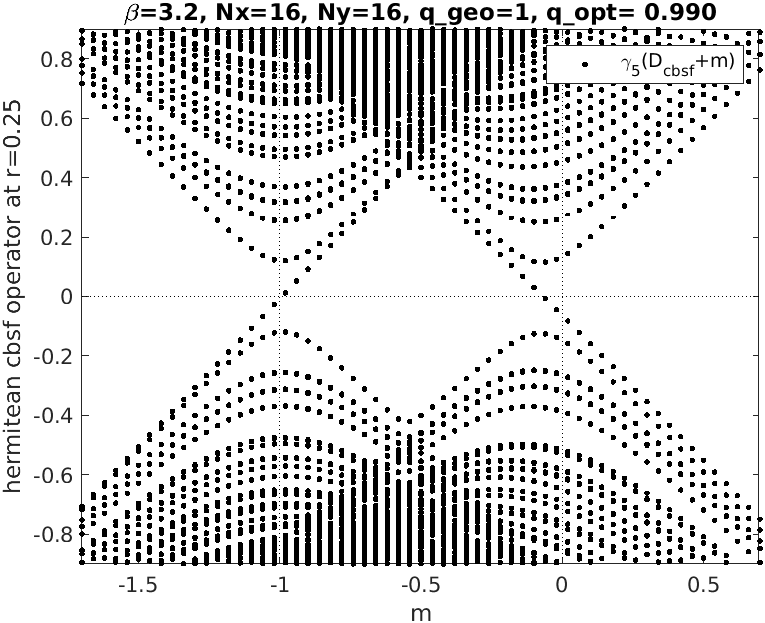}%
\vspace*{-2mm}
\caption{\label{fig:specflow_cbsf}\sl
Spectral flow of the ``central-branch-squared-and-flipped'' operator at $r\!=\!\frac{1}{4}$; eigen\-values of $\gaf(D_\mr{cbsf}+m)$ versus $m$. 
Relevant part near $m=0$ (left) and panoramic view (right).}
\end{figure}

In Fig.~\ref{fig:eigstem_cbsq} the eigenvalue spectrum of $D_\mr{cbs}$ is plotted.
In comparison to $D_\mr{cb}$ the recipe (\ref{def_cbs}, \ref{momrep_cbs}) squares the (horizontally acting) Wilson ``lifting term'' while the (vertically acting) ``derivative term'' is unaltered.
Hence the right (curved) branch in this figure is a superposition of what used to be the left-most and right-most branches of the Wilson operator,
while the left (near-straight) branch of $D_\mr{cbs}$ is more-or-less identical to the central branch of $D_\mr{cb}$.
Accordingly, the left (near-straight) branch of $D_\mr{cbs}$ hosts two wrong-chirality species, while the right (curved) branch hosts two right-chirality species.
A remarkable feature is the unusually small additive mass shift%
\footnote{This mass shift is due to the fact that the operators $C_\mu^2$ mix with the identity, while the operators $C_\mu C_\nu$ do not for $\mu\neq\nu$, cf.\ footnote \ref{foot:taste}.}
of the near-straight branch.
For each eigenvalue $\la_i$ we calculate the left-eigenvector $\<\ps_i|$ and the right-eigenvector $|\ps_i\>$ of $D_\mr{cbs}$ and determine the chirality $\<\ps_i|\gaf|\ps_i\>$.
The result is displayed as a needle at position $\la_i\in\mathbb{C}$.
We find two would-be zero-modes in the physical branch with chiralities close to $+1$, and two modes in the doubler branch, at $\mr{Re}(\la)\simeq4$, with chiralities close to $-1$.
In the physical branch nearby modes tend to have very small chiralities, while in the doubler branch adjacent modes are subject%
\footnote{This effect is mitigated by using a smaller $r$; for instance $r=\frac{1}{4}$ reduces the mixing significantly.
This is evident from comparing Fig.~\ref{fig:eigstem_cbsq} and Fig.~\ref{fig:eigstem_cbsf}.
The latter figure is \emph{not} a ``squeezed'' version (by a factor $\frac{1}{4}$) of the former one; the key difference is the amount of ``jumping'' of the needles near the needle pointing towards $-1$.}
to heavy mixing.

In Fig.~\ref{fig:specflow_cbsq} the spectral flow of the operator $H_\mr{cbs}=\gaf(D_\mr{cbs}+m)$ is displayed.
There is a two-fold (near-degenerate) up-crossing at $m\simeq0$, but there is no net crossing on a large scale.
The deviation of the up-crossing from $m=0$ is smaller than the deviation of the down-crossing from $m=-4r=-4$.
This matches, in the eigenvalue plot, the small offset of the physical branch from $\mr{Re}(\la)=0$ and the relatively large offset of the doubler branch from $\mr{Re}(\la)=4r=4$.

The ``central-branch-squared-and-flipped'' Dirac operator is defined as
\beq
D_\mr{cbsf}(x,y)=\sum_\mu \ga_\mu \nab_\mu(x,y)
+\frac{r}{a}d^2I_{x,y}
-\frac{r}{a}\Big[-\frac{a^2}{2} \sum_\mu \lap_\mu-dI\Big]_{x,y}^2
\label{def_cbsf}
\eeq
where the flipping operation is designed to interchange the physical and the right-most doubler branches.
In 2D this has no effect on the number of species, but in 4D it trades $6$ species for $2$.

In Fig.~\ref{fig:eigstem_cbsf} the eigenvalue spectrum of $D_\mr{cbsf}$ at $r=\frac{1}{4}$ is plotted.
If we were to stay with $r=1$ the figure would be a copy of Fig.~\ref{fig:eigstem_cbsq}, except for an inversion about $\mr{Re}(\la)=2$,
and the ``needle plot'' would result via the same operation from Fig.~\ref{fig:eigstem_cbsq}.
In this case we would be confronted with the unpleasant feature that the physical would-be zero-modes are in the middle of a region with heavy mixing, as evident from the little spikes nearby.
By choosing $r=\frac{1}{4}$ the mixings in the curved (now physical) branch are drastically reduced compared to Fig.~\ref{fig:eigstem_cbsq}.

In Fig.~\ref{fig:specflow_cbsf} the spectral flow of the operator $H_\mr{cbsf}=\gaf(D_\mr{cbsf}+m)$ is shown.
There is a two-fold (near-degenerate) down-crossing at $m\simeq0$, but there is no net crossing on a large scale.
The deviation of the down-crossing from $m=0$ is larger than the deviation of the up-crossing from $m=-4r=-1$.
This matches, in the eigenvalue plot, the large offset of the physical branch from $\mr{Re}(\la)=0$ and the much smaller offset of the doubler branch from $\mr{Re}(\la)=4r=1$.


\section{Karsten-Wilczek fermions\label{sec:kawi}}


The Karsten-Wilczek proposal is to restrict the Wilson term in (\ref{def_wils}) to the spatial components
\beq
\DKW(x,y)=\sum_\mu \ga_\mu \nab_\mu(x,y)
-\ri\frac{ra}{2}\ga_d \sum_{i=1}^{d-1} \lap_i(x,y)
\label{def_KW}
\eeq
with an extra factor $\ri\ga_d$ to make it anti-hermitean and anti-commuting with $\gaf$ \cite{Karsten:1981gd,Wilczek:1987kw}.
As a result, the Karsten-Wilczek (KW) operator is $\gaf$-hermitean, i.e.\ $\gaf \DKW \gaf=\DKW\dag$.
In the free-field limit the KW operator assumes a diagonal form in momentum space
\bea
\DKW(p)&=&\ri \sum_\mu \ga_\mu \frac{1}{a}\sin(ap_\mu)
+\ri\frac{r}{a}\ga_d \sum_{i=1}^{d-1} \{1-\cos(ap_i)\}
\nonumber
\\
&=&\ri \sum_\mu \ga_\mu \bar{p}_\mu
+\ri \frac{ra}{2}\ga_d \sum_{i=1}^{d-1} \hat{p}_i^2
\label{momrep_KW}
\eea
which again highlights the anti-hermitean nature of either term.

This formulation was shown to have $2$ species for $r=1$ in the original works \cite{Karsten:1981gd,Wilczek:1987kw}.
How this number decreases from $2^d$, at $r=0$, to $2$, at $r=1$, has been investigated in Ref.~\cite{Durr:2020yqa}.
In $d=4$ dimensions the number of species is reduced by $2,6,6$ at $r=1/6,1/4,1/2$, respectively, so the species chain is $16\to14\to8\to2$.
In $d=2$ dimensions the reduction takes place at $r=1/2$, so the species chain is $ 4\to2$.
Of course, the number of species is unchanged by a sign flip of $r$.
In Ref.~\cite{Durr:2020yqa} also the free-field (quark-level) dispersion relation of the KW operator is given.
In 2D the KW operator (\ref{def_KW}) takes the simple form (cf.\ App.~\ref{app:notation})
\beq
\DKW(x,y)=\sum_\mu \si_\mu \nab_\mu(x,y)
-\ri\frac{ra}{2}\si_2 \lap_1(x,y)
\;.
\label{twodim_KW}
\eeq

\begin{figure}[p]
\vspace*{-6mm}
\includegraphics[width=0.47\textwidth]{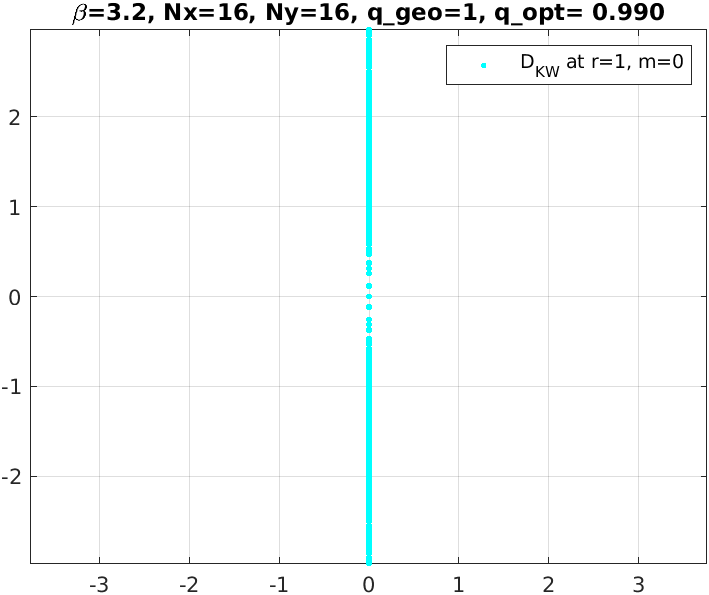}\hfill
\includegraphics[width=0.49\textwidth]{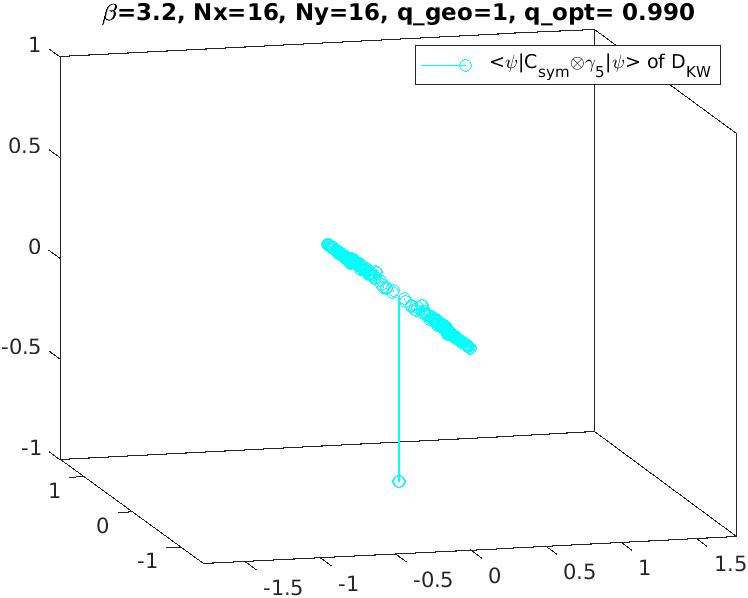}%
\vspace*{-2mm}
\caption{\label{fig:eigstem_kawi}\sl
Eigenvalues of the KW Dirac operator on a background with $q=1$ (left),
and ``needle plot'' of the $C_\mr{sym}\otimes\gaf$-chiralities in the pertinent left-right-eigenvector sandwich (right).
The standard $\gaf$-chiralities are exactly flat (not shown).}
\vspace*{+2mm}
\includegraphics[width=0.48\textwidth]{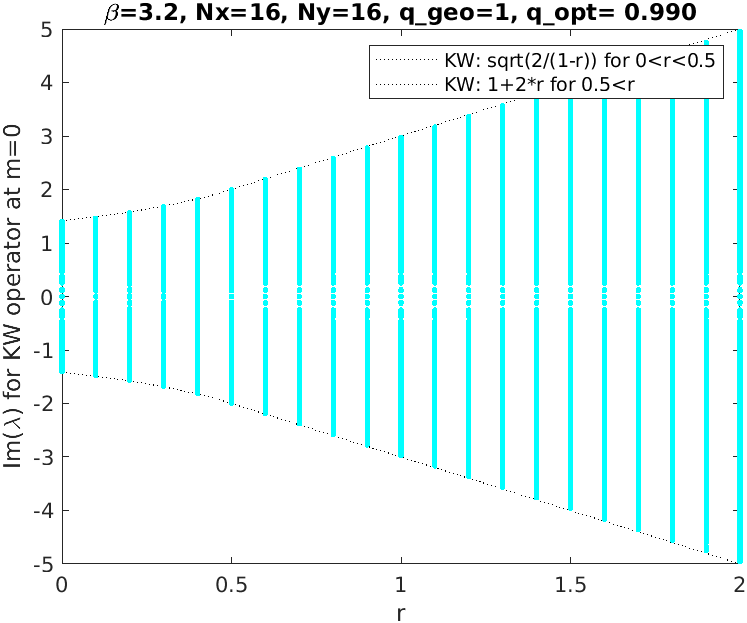}\hfill
\includegraphics[width=0.48\textwidth]{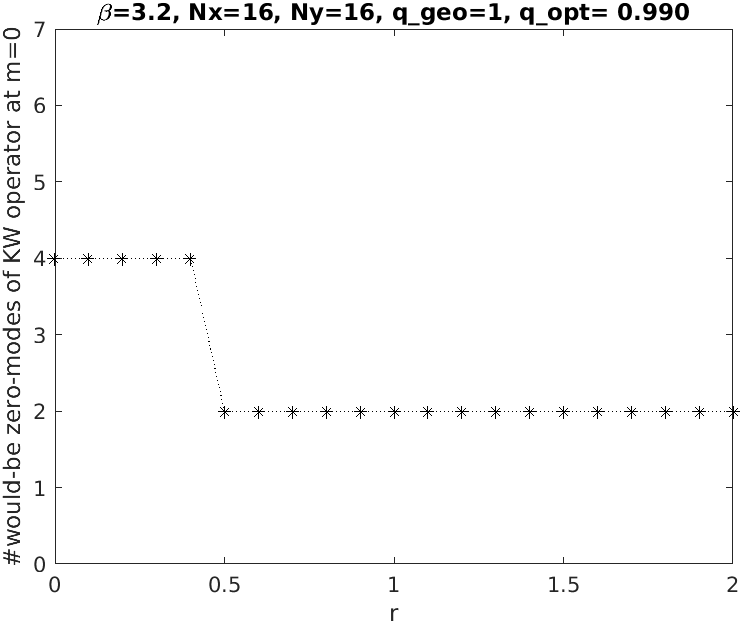}%
\vspace*{-4mm}
\caption{\label{fig:range_kawi}\sl
Eigenvalues of the KW Dirac operator on an interacting background as a function of the species-lifting parameter $r$ (left),
and number of would-be zero-modes versus $r$ (right).}
\vspace*{+2mm}
\includegraphics[width=0.48\textwidth]{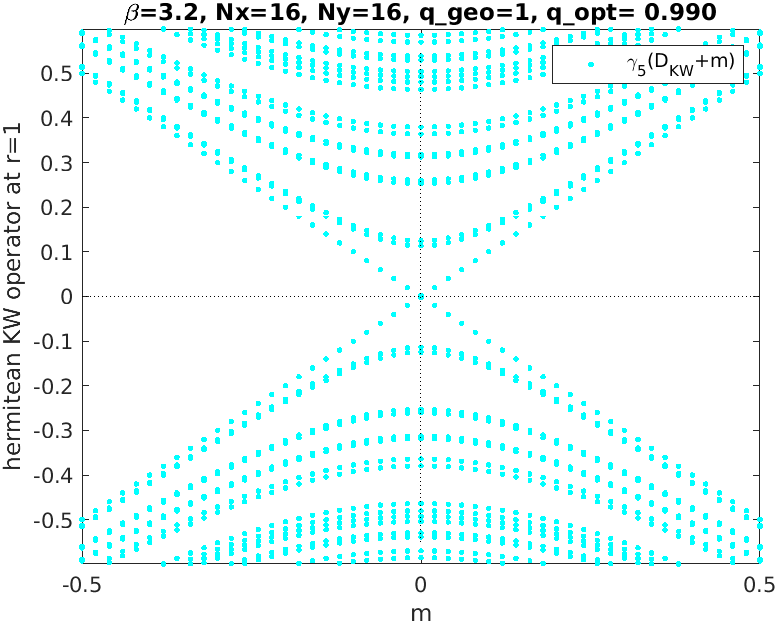}\hfill
\includegraphics[width=0.48\textwidth]{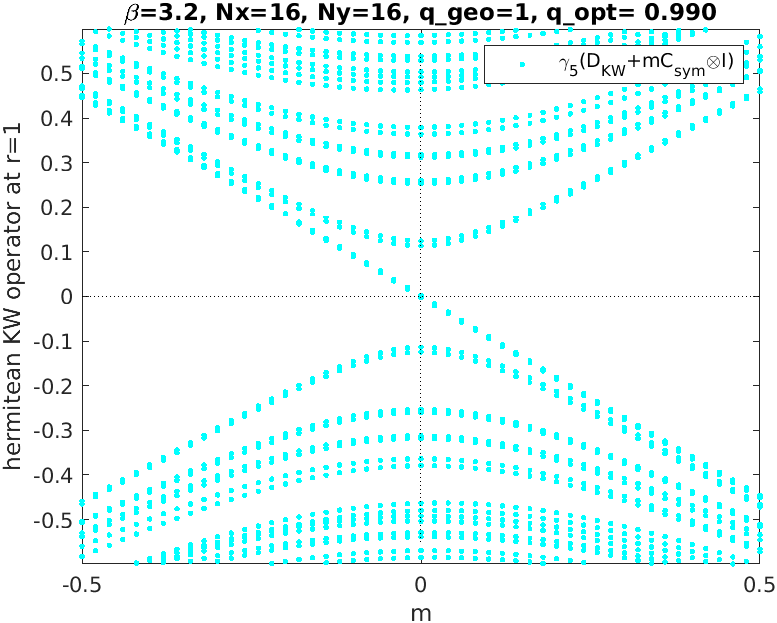}%
\vspace*{-4mm}
\caption{\label{fig:specflow_kawi}\sl
Spectral flow of the KW Dirac operator, i.e.\ eigenvalues of $\gaf(\DKW+m)$ versus $m$ (left),
and eigenvalues of $\gaf(\DKW+mC_\mr{sym}{\otimes}1)$ versus $m$ (right).}
\end{figure}

The eigenvalues of the KW operator (\ref{twodim_KW}) at $r=1$ on the same $q=1$ background as before are displayed in Fig.~\ref{fig:eigstem_kawi}.
The spectrum is purely imaginary, like in the staggered/naive case, but it stretches out to $\pm3$.
In the depleted part in the middle, eigenvalues come in near-degenerate pairs (the pair $\la=\pm0.00214\ri$ is represented by a single blob on the real axis).
We compute the right-eigenvector $|\ps_i\>$ and the left-eigenvector $\<\ps_i|$ for each eigenvalue $\la_i\in\mathbb{C}$.
If one were to choose the chirality operator $\gaf$, the result would be an entirely flat ``needle plot''.
Choosing an appropriate%
\footnote{\label{foot:KW_chiralities}
The KW operator has two zeros in the Brillouin zone, with opposite chiralities due to the Nielsen-Ninomiya theorem \cite{Karsten:1980wd,Nielsen:1981xu,Nielsen:1980rz}.
In the free field case they are located at $ap_d=0,\pi$ (with $ap_i=0$ for $i=1,...,d-1$), and encode the same chiralities as for naive fermions \cite{Durr:2020yqa}.
Therefore, the same chirality operators can be employed for KW and naive fermions.
We found good results with both $C_\mr{sym}\otimes\gaf$ and $[\frac{1}{2}(C_1+C_2)^2-1]\otimes\gaf$ at $r=1$.}
chirality operator like $C_\mr{sym}\otimes\gaf$ the situation is different;
for each mode the chirality $\<\ps_i|C_\mr{sym}\otimes\gaf|\ps_i\>$ is plotted as a needle at position $\la_i\in\mathbb{C}$.
One finds two needles reaching nearly down to $-1$, as expected for a two-species formulation and $q=1$.

In Ref.~\cite{Durr:2020yqa} we discuss, for $d=2$ and $d=4$, how the KW operator evolves from the naive operator as $r$ increases from $0$ to $1$.
In particular we derive the spectral bound $|\mr{Im}(\la_\mr{KW})|\leq\sqrt{2/(1-r)}$ for $0\leq r\leq\frac{1}{2}$, and $|\mr{Im}(\la_\mr{KW})|\leq1+2r$ for $\frac{1}{2}\leq r$,
both valid in the free-field case in 2D.
In Fig.~\ref{fig:range_kawi} the eigenvalue spectrum is shown for a number or $r$-values, along with the spectral bound mentioned.
The free-field bound seems to give a rather accurate estimate of the actual spectral range in the interacting case.
Defining a would be zero-mode by the (somewhat arbitrary) criterion $|\mr{Im}(\la_\mr{KW})|<1/\sqrt{2 N_x N_y}$,
the number of would-be zero-modes is seen to drop from $4$ to $2$ in the vicinity of $r=1/2$, in line with expectations \cite{Durr:2020yqa}.

The spectral flow plot with the inappropriate choice of chirality operator, i.e.\ the eigenvalues of $\gaf(\DKW+m)$ versus $m$, is shown in Fig.~\ref{fig:specflow_kawi}.
As expected, there is no crossing in the vicinity of $m=0$.
The situation is different with an appropriate chirality operator; the eigenvalues of $\gaf(\DKW+mC_\mr{sym}{\otimes}1)$ show a two-fold down-crossing near $m=0$.
At a superficial level, the latter plot looks similar to the ``good'' staggered and naive plots in Figs.~\ref{fig:specflow_stag}, \ref{fig:specflow_naiv}, respectively.
Still, there are two notable differences.
Compared to the ``good'' staggered plot the \emph{intra-taste splitting is smaller} (cf.\ discussion in Sec.~\ref{sec:bocr}).
Compared to the ``good'' naive plot, in addition the (exact) two-fold degeneracy is missing.


\section{Borici-Creutz fermions\label{sec:bocr}}


The basis for Borici-Creutz fermions in $d$ space-time dimensions is the idempotent operator
\beq
\Gamma=\frac{1}{\sqrt{d}}\sum_\mu\ga_\mu
\quad \mbox{with} \quad
\Gamma^2
=\frac{1}{2d}\{\sum_\al\ga_\al,\sum_\be\ga_\be\}
=\frac{2d}{2d}=1
\label{def_big}
\eeq
and $\{\Gamma,\ga_\mu\}=\frac{2}{\sqrt{d}}$ and $\{\Gamma,\gaf\}=0$.
This suggests to define the dual gamma-matrices
\beq
\ga_\mu'=\Gamma \ga_\mu \Gamma
=\Big( \frac{2}{\sqrt{d}} - \ga_\mu \Gamma \Big)\Gamma
=\frac{2}{\sqrt{d}}\Gamma-\ga_\mu
\label{def_dualgamma}
\eeq
which are hermitean and satisfy the Dirac-Clifford algebra, since (\ref{def_dualgamma}) implies $\{\ga_\mu',\ga_\nu'\}=2\de_{\mu\nu}$ and $\{\Gamma,\ga_\mu'\}=\frac{2}{\sqrt{d}}$.
Furthermore, one finds $\{\ga_\mu,\ga_\nu'\}=\frac{4}{d}-2\de_{\mu\nu}=\{\ga_\mu',\ga_\nu\}$.

The Borici-Creutz (BC) proposal is to dress the Wilson term in (\ref{def_wils}) with $\ri$ times (\ref{def_dualgamma})
\beq
\DBC(x,y)=\sum_\mu \ga_\mu \nab_\mu(x,y)
-\ri\frac{ra}{2}\sum_\mu \ga_\mu' \lap_\mu(x,y)
\label{def_BC}
\eeq
where our second term differs in sign from the original proposal \cite{Creutz:2007af,Borici:2007kz}.
Note that the second term is anti-hermitean and anti-commutes with $\gaf$, since
\beq
\ga_\mu'\gaf=\Gamma\ga_\mu\Gamma\gaf=-\Gamma\ga_\mu\gaf\Gamma=\Gamma\gaf\ga_\mu\Gamma=-\gaf\Gamma\ga_\mu\Gamma=-\gaf\ga_\mu'
\eeq
and this renders the BC operator $\gaf$-hermitean, i.e.\ $\gaf \DBC \gaf=\DBC\dag$.
In the free-field limit the BC operator assumes a diagonal form in momentum space
\bea
\DBC(p)&=&\ri \sum_\mu \ga_\mu \bar{p}_\mu
+\ri\frac{r}{a}\sum_\mu \ga_\mu' \{1-\cos(ap_\mu)\}
\nonumber
\\
&=&\ri \sum_\mu \ga_\mu \bar{p}_\mu
+\ri \frac{ra}{2} \sum_\mu \ga_\mu' \hat{p}_\mu^2
\label{momrep_BC}
\eea
in which the bracket $\{1-\cos(ap_\mu)\}$ may be split and the sum over $\ga_\mu'$ performed by means of
\beq
\sum_\mu \ga_\mu'=2\sqrt{d}\Gamma-\sum_\mu\ga_\mu
=2\sqrt{d}\Gamma-\sqrt{d}\Gamma=\sqrt{d}\Gamma
\;.
\eeq
The free-field form (\ref{momrep_BC}) highlights the invariance under any permutation of the $d$ axes.

This formulation was shown to have $2$ species for $r=1$ in the original works \cite{Creutz:2007af,Borici:2007kz}.
How this number decreases from $2^d$, at $r=0$, to $2$, at $r=1$, has been investigated in Ref.~\cite{Durr:2020yqa}.
In 4D one starts with $16$ species, and this number decreases by $6$ at $r=1/\sqrt{3}$, and by $8$ at $r=1/\sqrt{2}$; so the species chain is $16\to10\to2$.
In 2D one starts with $4$ species, and this number decreases by $2$ at $r=1/\sqrt{3}$; so the species chain is $4\to2$.
Of course, the number of species is unchanged by a sign flip of $r$.
In Ref.~\cite{Durr:2020yqa} also the free-field (quark-level) dispersion relation of the BC operator is given.
In 2D the BC operator (\ref{def_BC}) takes the simple form (cf.\ App.~\ref{app:notation})
\beq
\DBC(x,y)=\sum_\mu \si_\mu \nab_\mu(x,y)
-\ri\frac{ra}{2} \si_2 \lap_1(x,y)
-\ri\frac{ra}{2} \si_1 \lap_2(x,y)
\label{twodim_BC}
\eeq
and comparing this to (\ref{twodim_KW}) shows that the BC operator is not a symmetrized form of the KW operator; it has an extra term.
This is why, in Eqns.~(\ref{def_BC}, \ref{momrep_BC}), the sign of the $r$-dependent term differs from the literature.
With our convention the joint terms in Eqns.~(\ref{twodim_KW}, \ref{twodim_BC}) have like sign.

\begin{figure}[p]
\vspace*{-6mm}
\includegraphics[width=0.47\textwidth]{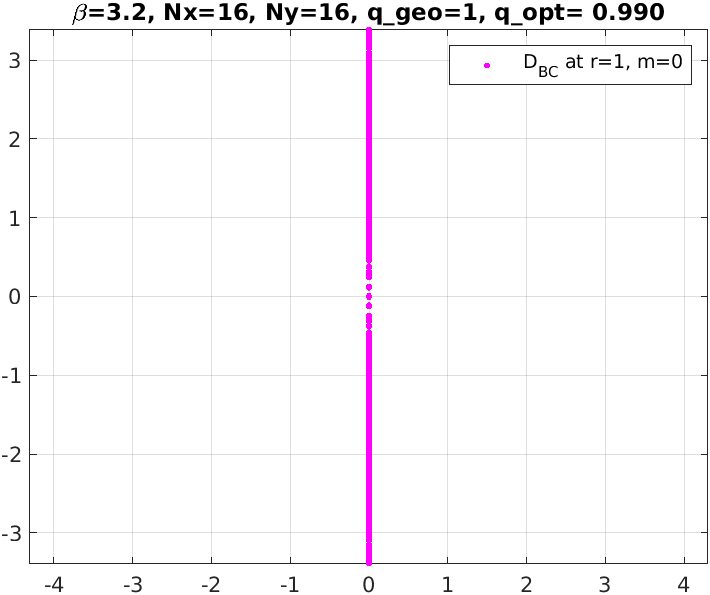}\hfill
\includegraphics[width=0.49\textwidth]{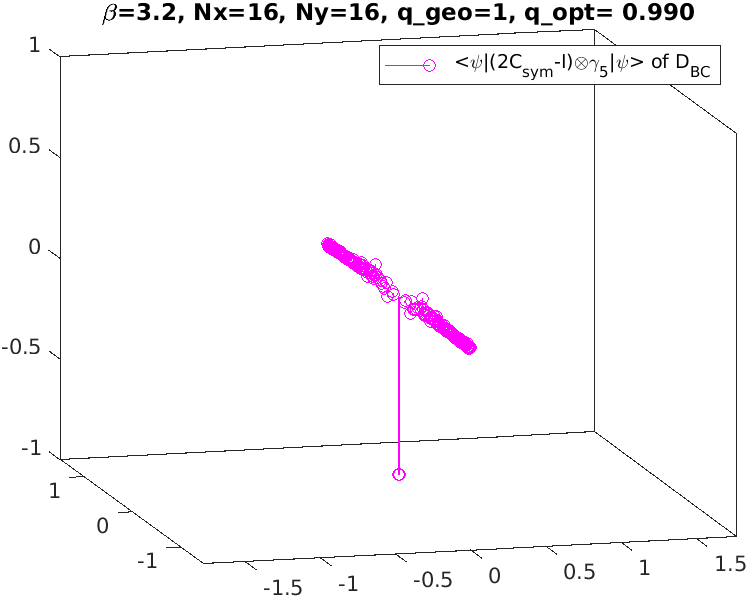}%
\vspace*{-2mm}
\caption{\label{fig:eigstem_bocr}\sl
Eigenvalues of the BC Dirac operator on a background with $q=1$ (left),
and ``needle plot'' of the $[2C_\mr{sym}-1]\otimes\gaf$-chiralities in the pertinent left-right-eigenvector sandwich (right).
The standard $\gaf$-chiralities are exactly flat (not shown).}
\vspace*{+2mm}
\includegraphics[width=0.48\textwidth]{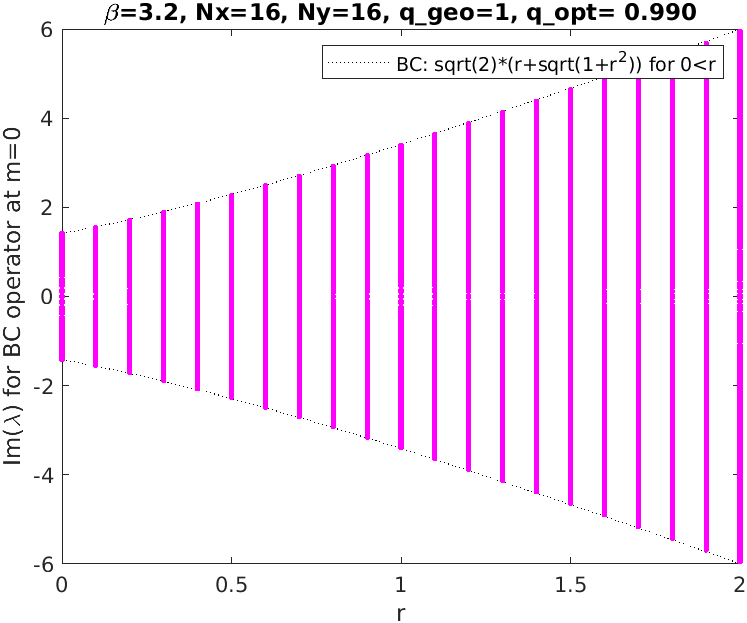}\hfill
\includegraphics[width=0.48\textwidth]{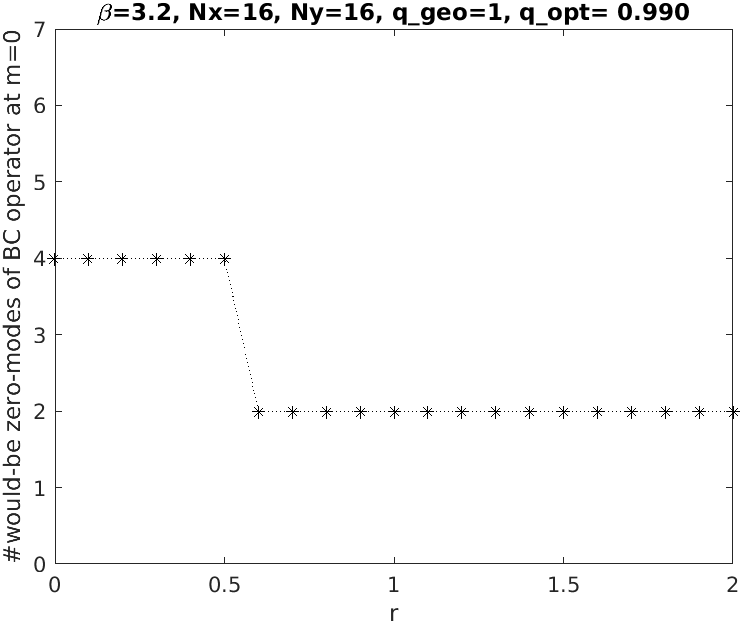}%
\vspace*{-4mm}
\caption{\label{fig:range_bocr}\sl
Eigenvalues of the BC Dirac operator on an interacting background as a function of the species-lifting parameter $r$ (left),
and number of would-be zero-modes versus $r$ (right).}
\vspace*{+2mm}
\includegraphics[width=0.48\textwidth]{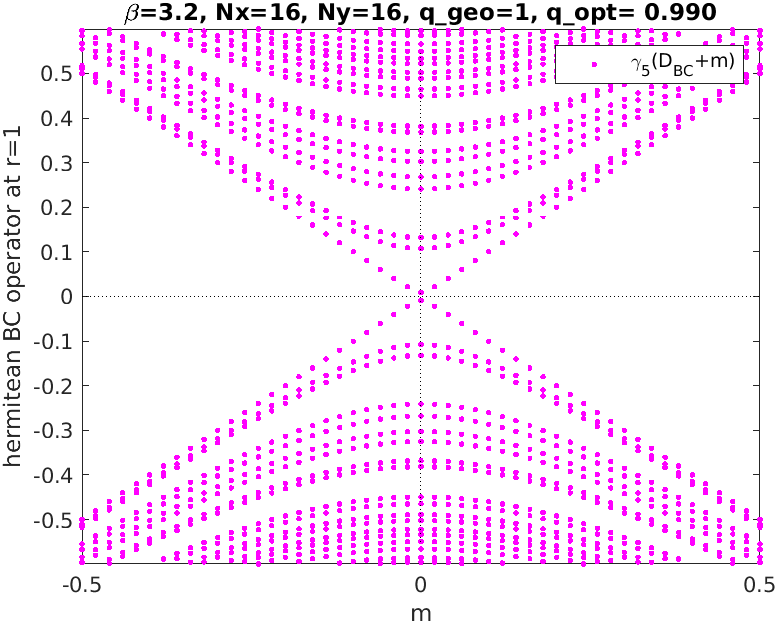}\hfill
\includegraphics[width=0.48\textwidth]{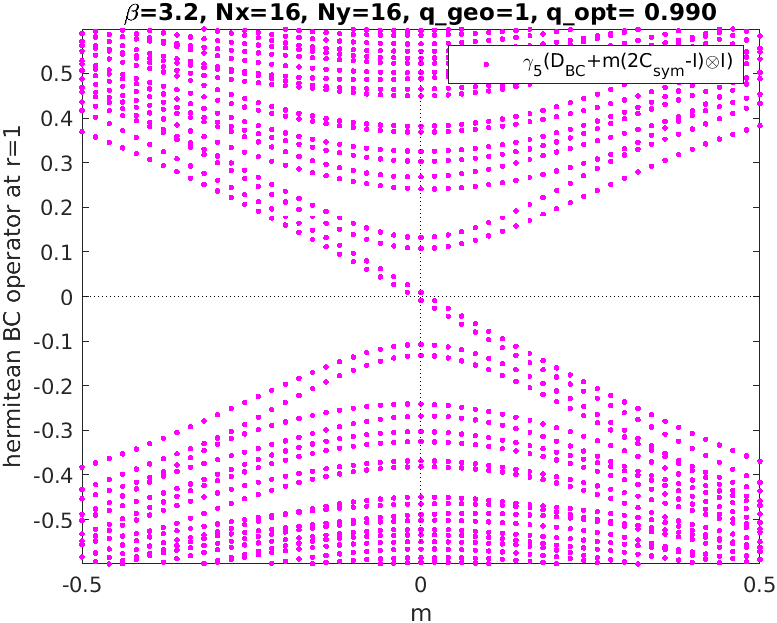}%
\vspace*{-4mm}
\caption{\label{fig:specflow_bocr}\sl
Spectral flow of the BC Dirac operator, i.e.\ eigenvalues of $\gaf(\DBC+m)$ versus $m$ (left),
and eigenvalues of $\gaf(\DBC+m[2C_\mr{sym}-1]{\otimes}1)$ versus $m$ (right).}
\end{figure}

The eigenvalues of the BC operator (\ref{twodim_BC}) at $r=1$ on the same $q=1$ background as before are displayed in Fig.~\ref{fig:eigstem_bocr}.
The spectrum is purely imaginary (as for the previously discussed chiral actions), but this time it stretches out to $\pm(2+\sqrt{2})$.
In the depleted part in the middle, eigenvalues come in near-degenerate pairs (the pair $\la=\pm0.00854\ri$ is represented by a single blob on the real axis).
We compute the right-eigenvector $|\ps_i\>$ and the left-eigenvector $\<\ps_i|$ for each eigenvalue $\la_i\in\mathbb{C}$.
With $\gaf$ as chirality operator, one obtains $\<\ps_i|\gaf|\ps_i\>=0$ for all modes $\ps_i$.
Choosing $[2C_\mr{sym}-1]\otimes\gaf$ as chirality operator%
\footnote{\label{foot:BC_chiralities}
As for KW fermions, the two surviving zero-modes of BC fermions must encode opposite $\gaf$ chiralities, due to the Nielsen-Ninomiya theorem \cite{Karsten:1980wd,Nielsen:1981xu,Nielsen:1980rz}.
An extended splitting operator separates these cleanly if it assumes opposite (non-zero) real values at these two positions.
The two zero-modes of BC fermions surviving in the free theory are located at $ap_\mu=0$ and $ap_\mu=\kappa(r)$,
with $\kappa(r)\equiv-2\arctan(1/r)$ for all $d$ components of $ap$~\cite{Durr:2020yqa}.
For the canonical value $r=1$, the second mode is thus at $\kappa(1)=-\pi/2$.
While the operator $C_\mr{sym}(k)$ assumes the value $1$ at $ap_\mu=0$, it yields $0$ at $ap_\mu=-\pi/2$.
A deformation of $C_\mr{sym}(k)$ that realizes the desired sign change is $(1+A(r))C_{\mathrm{sym}}-A(r)$ with
$A(r)=[2+\cos^2(\kappa(r))]/[2-\cos^2(\kappa(r))]$; this yields $A(1)=1$.
We found good chirality results with both $[2C_\mr{sym}-1]\otimes\gaf$ and $[\frac{1}{2}(C_1+C_2)^2-1]\otimes\gaf$ at $r=1$.}
the situation is different; for each mode the chirality $\<\ps_i|[2C_\mr{sym}-1]\otimes\gaf|\ps_i\>$ is plotted as a needle at position $\la_i\in\mathbb{C}$.
One finds two needles reaching nearly down to $-1$, as expected for a two-species formulation and $q=1$.

In Ref.~\cite{Durr:2020yqa} we discuss how the BC operator evolves from the naive operator as $r$ increases from $0$ to $1$.
In particular we derive the spectral bound $|\mr{Im}(\la_\mr{BC})|\leq\sqrt{d}(r+\sqrt{1+r^2})$ for $0\leq r$, valid in the free-field case in $d$ dimensions.
In Fig.~\ref{fig:range_bocr} the eigenvalue spectrum is shown for a number or $r$-values, along with the spectral bound mentioned (for $d=2$).
The free-field bound seems to give a rather accurate estimate of the actual spectral range in the interacting case.
Defining a would be zero-mode by the same criterion as in Sec.~\ref{sec:kawi},
their number is seen to evolve from $4$ to $2$ in the vicinity of $r=1/\sqrt{3}$, in line with expectations \cite{Durr:2020yqa}.

The spectral flow plot with the inappropriate choice of chirality operator, i.e.\ the eigenvalues of $\gaf(\DBC+m)$ versus $m$, is shown in Fig.~\ref{fig:specflow_kawi}.
As expected, there is no crossing in the vicinity of $m=0$.
The situation is different with the appropriate chirality operator; the eigenvalues of $\gaf(\DBC+m[2C_\mr{sym}-1]{\otimes}1)$ show a two-fold down-crossing near $m=0$.
At a superficial level, the latter plot looks similar to the ``good'' staggered plot in Fig.~\ref{fig:specflow_stag}, the ``good'' naive plot in Fig.~\ref{fig:specflow_naiv},
and the ``good'' KW plot in Fig.~\ref{fig:specflow_kawi} (modulo degeneracies).

\begin{figure}[!tb]
\includegraphics[width=0.49\textwidth]{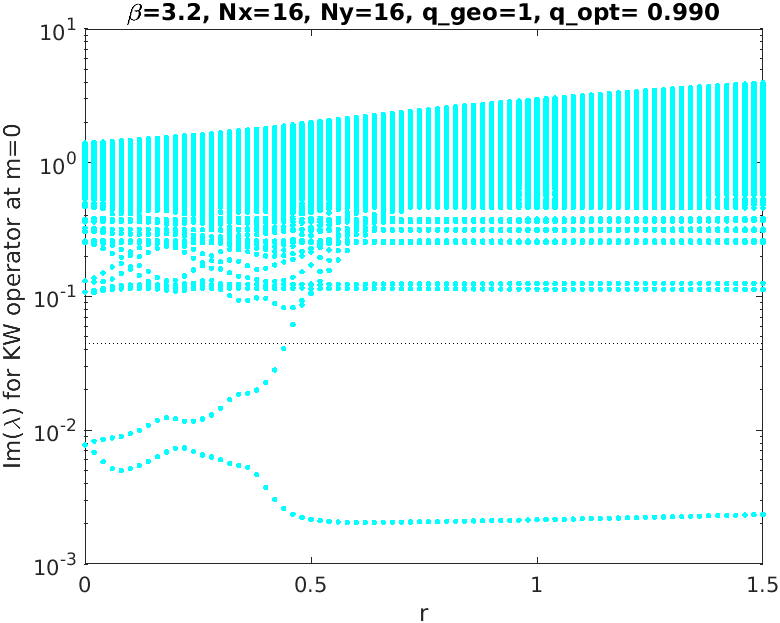}\hfill
\includegraphics[width=0.49\textwidth]{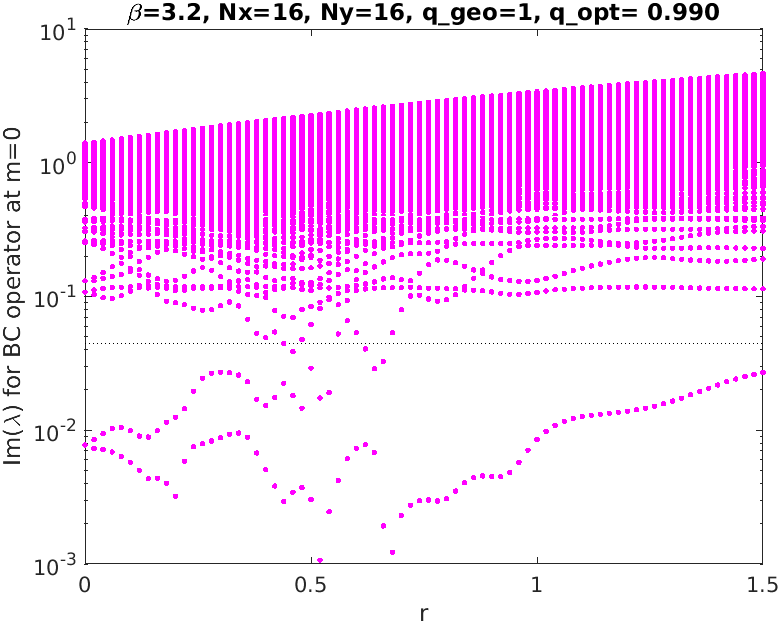}%
\vspace*{-2mm}
\caption{\label{fig:logspectrum}\sl
Upper half of Figs.~\ref{fig:range_kawi}, \ref{fig:range_bocr} (left), but with logarithmic $y$-scale.
The threshold for an eigenvalue to be considered a ``would-be zero-eigenvalue'' is indicated by a dotted line.}
\end{figure}

Upon comparing these four plots more diligently, one notices that (apart from the extra two-fold exact degeneracy in the naive case)
these plots differ by the \emph{size of the taste breaking}.
Taking the splitting between the two down-crossings in Figs.~\ref{fig:specflow_stag} and \ref{fig:specflow_naiv} as a basis,
the splitting in the KW case (Fig.~\ref{fig:specflow_kawi}) seems smaller, while in the BC case (Fig.~\ref{fig:specflow_bocr}) it seems comparable or larger.
To corroborate this finding we display in Fig.~\ref{fig:logspectrum} how the (positive) eigenvalues $\la_\mr{KW}/\ri$ and $\la_\mr{BC}/\ri$ evolve from the naive ones,
as $r$ grows from $0$ to $1.5$, this time with a logarithmic scale on the ordinate.
The two-fold degeneracy of the naive action is split for any $r>0$, but the behavior of the (one) remaining would-be zero-eigenvalue is different in the two panels.
In the KW case it behaves very smoothly, and at $r\simeq1$ it is \emph{smaller} than in the naive/staggered case.
On the other hand, in the BC case the would-be zero-eigenvalue performs wild movements in the vicinity of the pole-merger zone at $r=1/\sqrt{3}$ (see Ref.~\cite{Durr:2020yqa}
for details), and at $r\simeq1$ the deviation of the would-be zero-eigenvalue from zero is \emph{comparable in size} to the naive/staggered case.
This raises further questions \cite{Bedaque:2008xs,Kimura:2011ik,Misumi:2012uu,Misumi:2012ky,Weber:2016dgo}; we shall briefly comment on this in Sec.~\ref{sec:conclusions}.


\section{KW and BC fermions with species-lifting terms\label{sec:KWBC}}


In Secs.~\ref{sec:stagadam}, \ref{sec:naiv}, \ref{sec:cebr}, \ref{sec:kawi} and \ref{sec:bocr} we learned about the close relationship between a ``good'' chirality operator $X$ which measures the chiralities $\<\ps_i|X|\ps_i\>$
of the eigenmodes $\ps_i$ of a given (doubled) $D$ and the operator $D+r(1\pm X\otimes\ep)$ or $D+r(1\pm X\otimes\gaf)$ in which the tastes are separated%
\footnote{Starting from the staggered Operator with $2^{d/2}$ species in $d$ dimensions, the resulting Adams operator (\ref{def_adam})
has $2^{d/2-1}$ left-handed species in one branch and an equal number of right-handed species in the other branch.
And starting from the naive operator with $2^d$ species, the resulting Adams-like operator (\ref{def_like}) has $2^{d-1}$ left-handed species
in one branch and an equal number of right-handed species in the other branch.}
according to their chiralities.
For KW fermions both $X=  C_\mr{sym}   \otimes\gaf$ and $X=[\frac{1}{2}(C_1+C_2)^2-1]\otimes\gaf$ turned out to be good chirality operators, see footnote~\ref{foot:KW_chiralities}.
For BC fermions both $X=[2C_\mr{sym}-1]\otimes\gaf$ and $X=[\frac{1}{2}(C_1+C_2)^2-1]\otimes\gaf$ were found to be good chirality operators, see footnote~\ref{foot:BC_chiralities}.
Hence the question arises whether one may add such an operator (without the factor $\gaf$) to separate the two species (``tastes'') present in $\DKW$ or $\DBC$.

\begin{figure}[!tb]
\includegraphics[width=0.48\textwidth]{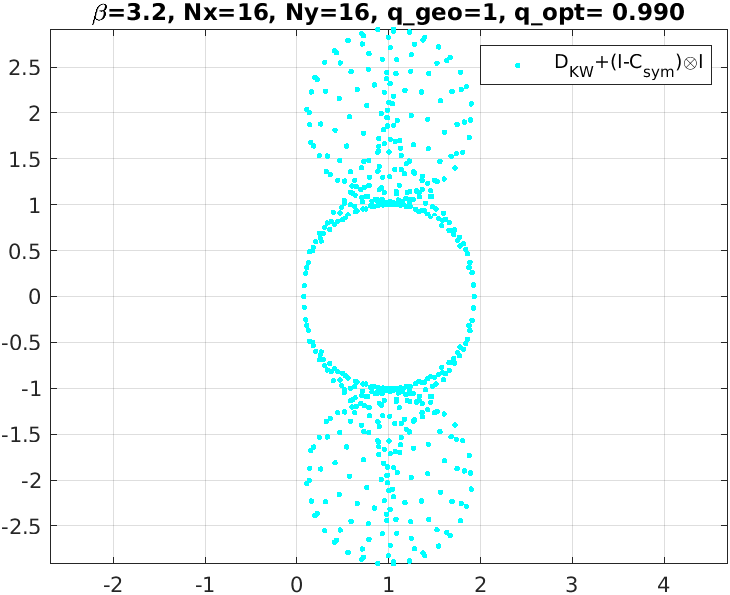}\hfill
\includegraphics[width=0.48\textwidth]{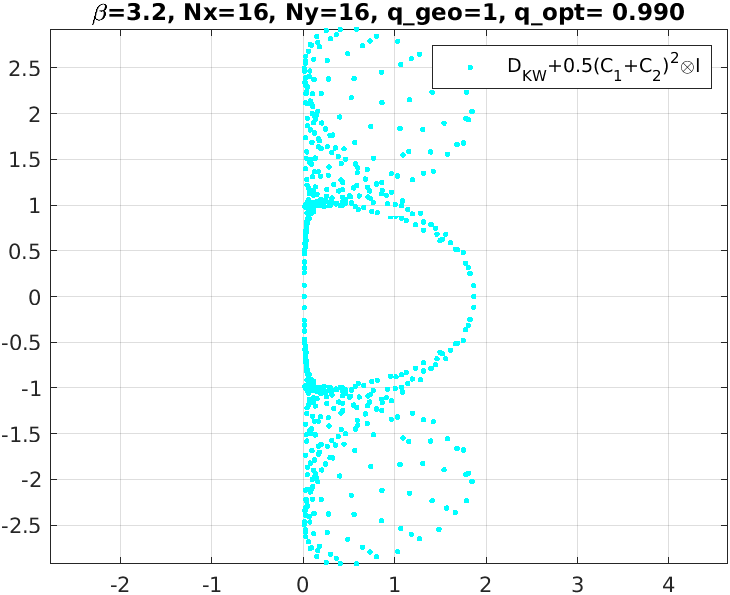}\\[2mm]
\includegraphics[width=0.49\textwidth]{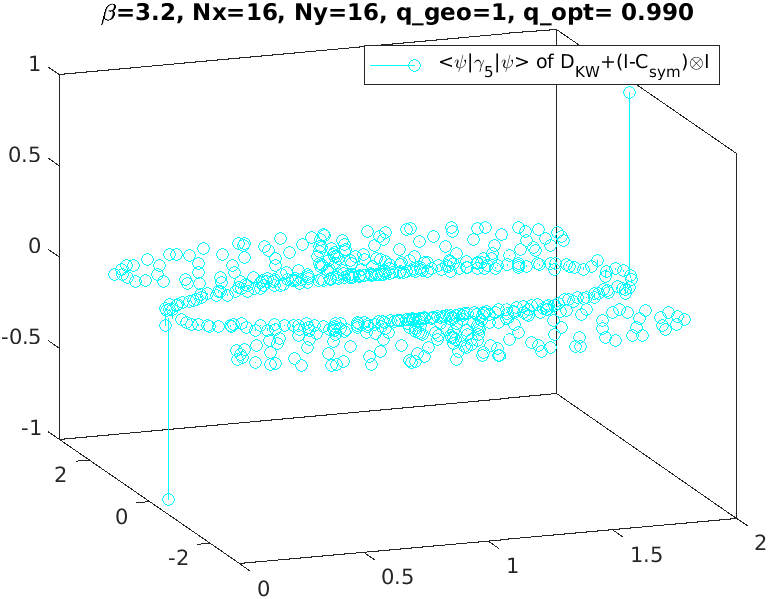}\hfill
\includegraphics[width=0.49\textwidth]{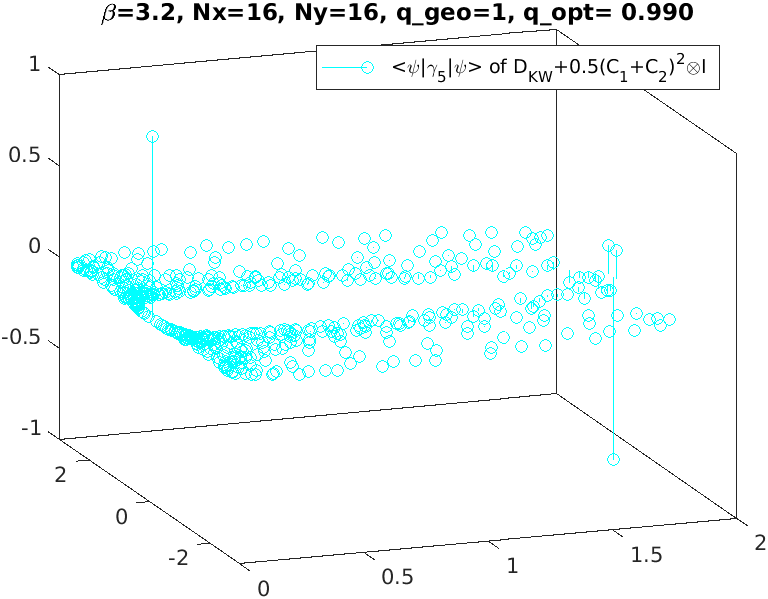}%
\vspace*{-2mm}
\caption{\label{fig:eigstem_kawi_modified}\sl
Eigenvalues of $\DKW$ with two lifting terms on a background with $q=1$, and
``needle plots'' of the $\gaf$-chiralities in the pertinent left-right-eigenvector sandwiches.}
\end{figure}

\begin{figure}[!tb]
\includegraphics[width=0.48\textwidth]{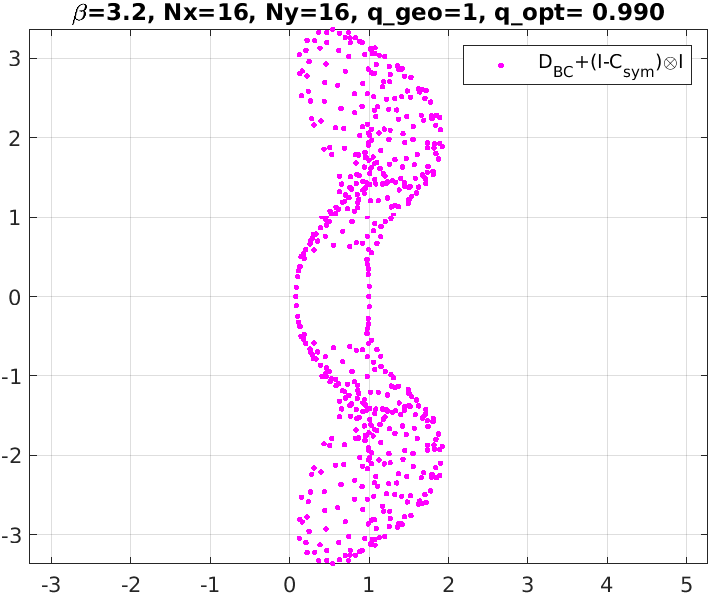}\hfill
\includegraphics[width=0.48\textwidth]{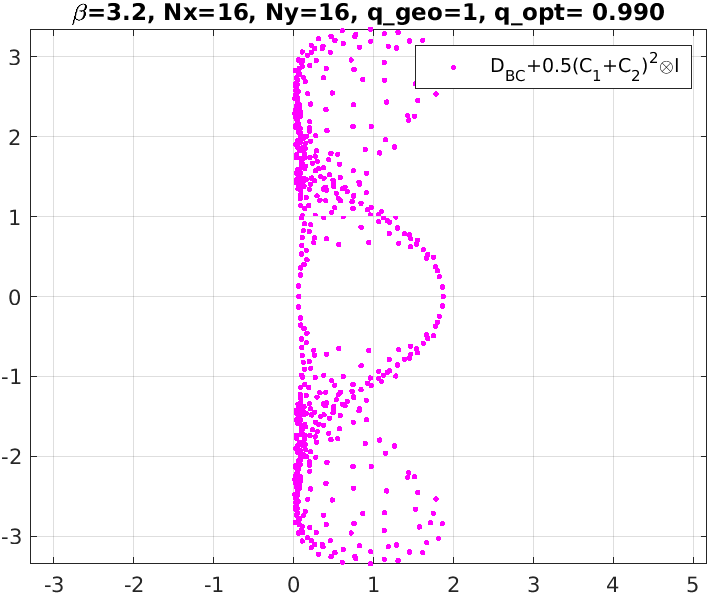}\\[2mm]
\includegraphics[width=0.49\textwidth]{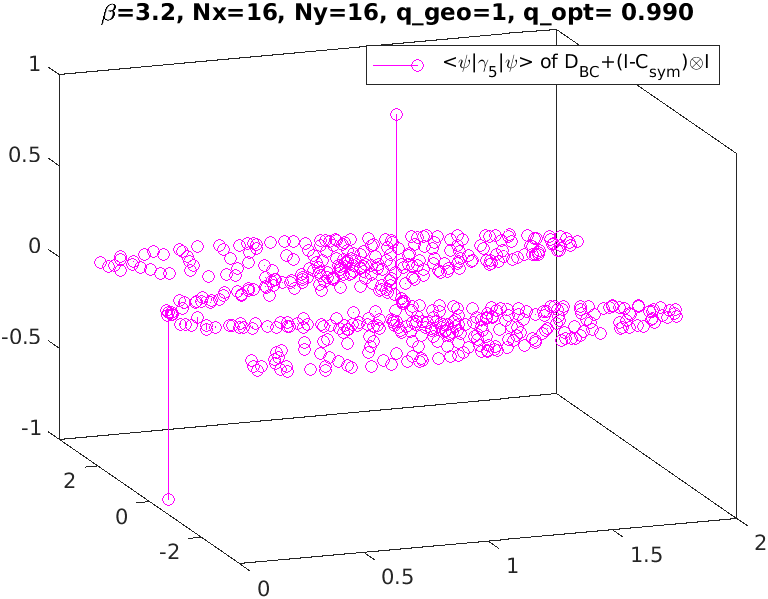}%
\includegraphics[width=0.49\textwidth]{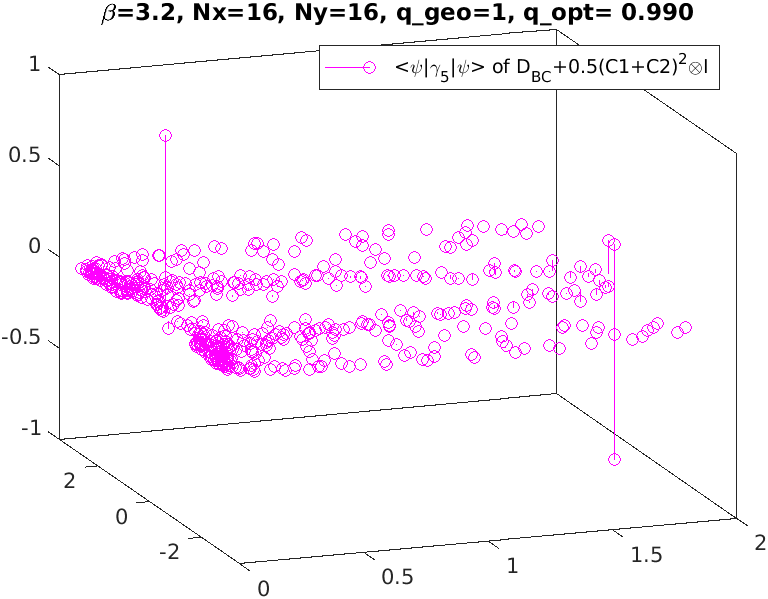}%
\vspace*{-2mm}
\caption{\label{fig:eigstem_bocr_modified}\sl
Eigenvalues of $\DBC$ with two lifting terms on a background with $q=1$, and
``needle plots'' of the $\gaf$-chiralities in the pertinent left-right-eigenvector sandwiches.}
\end{figure}

In Fig.~\ref{fig:eigstem_kawi_modified} the eigenvalues of the operators $\DKW+s(1-C_\mr{sym})\otimes1$
and $\DKW+\frac{s}{2}(C_1+C_2)^2\otimes1$ at $s=1$ are shown (with $r=1$ in $\DKW$).
Both species-lifting terms work fine, but the curvature of the curved branches%
\footnote{For the operator $\DKW+(1-C_\mr{sym})\otimes1$ the eigenvalues in the left branch resemble the eigenvalues in the physical branch of $\DW$,
and the right branch resembles the rightmost branch of $\DW-2$.
On the other hand in $\DKW+\frac{s}{2}(C_1+C_2)^2\otimes1$ it takes $s=\frac{1}{2}$ to make the right branch mimic the rightmost branch of $\DW-3$.}
agrees only in two of three cases with the curvature of the appropriate branch of the Wilson operator.
The panels with the respective $\gaf$-chiralities show that in either branch there is exactly one mode%
\footnote{\label{foot:sign}
In the left panel of Fig.~\ref{fig:eigstem_kawi_modified} the needle near $\la=0$ points downwards, so everything is fine.
In the right panel this needle points upwards, so consistency with the remainder of this article is lost.
This could be avoided by using the operator $\DKW+[2-\frac{1}{2}(C_1+C_2)^2]\otimes1$ instead.}
with chirality close to $\pm1$.
This is different from the situation encountered in Sec.~\ref{sec:kawi}, where the chiralities of $\DKW$ needed to be measured with $C_\mr{sym}\otimes\gaf$ or $\frac{1}{2}(C_1+C_2)^2\otimes\gaf$.
But this difference is completely analogous to the difference between the $\DS$ and $\DA$ in Sec.~\ref{sec:stagadam} or the difference between $\DN$ and $D_\mr{like}$ in Sec.~\ref{sec:naiv}.

In Fig.~\ref{fig:eigstem_bocr_modified} the eigenvalues of the operators $\DBC+s(1-C_\mr{sym})\otimes1$
and $\DBC+\frac{s}{2}(C_1+C_2)^2\otimes1$ at $s=1$ are shown (with $r=1$ in $\DBC$).
Both species-lifting terms work fine, but the curvature of the curved branches%
\footnote{The eigenvalues in the curved branch of $\DBC+(1-C_\mr{sym})\otimes1$ approximately coincide with the eigenvalues in the physical branch of $\DW$.
On the other hand in $\DBC+\frac{s}{2}(C_1+C_2)^2\otimes1$ it takes $s=\frac{1}{2}$ to ensure that the eigenvalues in the right branch match those in the rightmost branch of $\DW-3$.}
agrees only in one of two cases with the curvature of the appropriate branch of the Wilson operator.
The panels with the respective $\gaf$-chiralities show that in either branch there is exactly one mode%
\footnote{As to the signs of the needles footnote~\ref{foot:sign} applies again.}
with chirality close to $\pm1$.
Again, either construction repeats the reasoning which led to the Adams operator (\ref{def_adam}) or the Adams-like operator (\ref{def_like}).

In summary both $\DKW$ and $\DBC$ may be equipped with a species-lifting term.
All four options discussed yield an undoubled fermion operator with additive mass renormalization.
In the event $s(1-C_\mr{sym})\otimes1$ is added, we recommend using $s=1$.
In the event $\frac{s}{2}(C_1+C_2)^2\otimes1$ is added, we recommend using $s=\frac{1}{2}$, as this reduces unwanted mixings in the unphysical branch.






\section{Conclusions\label{sec:conclusions}}


Our goal was to provide evidence in the interacting theory that two minimally doubled fermion actions,
namely Karsten-Wilczek \cite{Karsten:1981gd,Wilczek:1987kw} and Borici-Creutz \cite{Creutz:2007af,Borici:2007kz} fermions,
perceive a global topological charge as foreseen in the seminal ``anomaly'' paper by Karsten and Smit \cite{Karsten:1980wd}.

In case of non-minimally doubled actions with (remnant exact) chiral symmetry, i.e.\ for staggered and naive fermions, it is known (perhaps not widely so)
that special diligence is needed to select an appropriate chirality operator $X$ to see the needles in the chirality plots $\<\ps_i|X|\ps_i\>$ at position $\la_i\in\mathbb{C}$.
In case of Karsten-Wilczek fermions choosing $X$ as $C_\mr{sym}\otimes\gaf$ or $\frac{1}{2}[(C_1+C_2)^2-1]\otimes\gaf$ yields good results,
and with Borici-Creutz fermions a similar statement holds true for $X$ being $[2C_\mr{sym}-1]\otimes\gaf$ or $\frac{1}{2}[(C_1+C_2)^2-1]\otimes\gaf$.

We find that any appropriate choice of $X$ for the ``needle plot'' would always yield the expected number of crossings in the spectral flow plot,
and it would result in a useful definition of the topological charge via the ``trace formula'' discussed in App.~\ref{app:charges} and App.~\ref{app:analytic}.

To stress the universality of the underlying concept, we opted for showing similar plots with Wilson, Brillouin, staggered, Adams, naive and Adams-like fermions,
plus two more varieties dubbed ``central-branch-squared'' and ``central-branch-squared-and-flipped'' fermions (of which the former one has very small additive mass shift).
To limit the overall length, our numerics was restricted to 2D, but we plan to show similar plots in 4D at some point in the future.

An important part of the discussion focused on the intimate relation between a working chirality operator $X$ for a given fermion action $D$
and the lifting term that is needed to separate branches in $D$ \emph{according to their chirality} (typically $X\otimes\gaf$ or $X\otimes\ep$).
This viewpoint emphasizes that Adams fermions are derived from staggered fermions in essentially the same way as the the Adams-like action (\ref{def_like}) is derived from the naive action (\ref{def_naiv}).
And it suggests dedicated splitting terms by means of which one of the species sitting in $\DKW$ or $\DBC$ can be lifted to become a doubler mode
(albeit at the price of loosing the remnant chiral symmetry).

Finally the smallness of the (one) would-be zero-eigenvalue of $\DKW$ at $r=1$ in Fig.~\ref{fig:logspectrum} provides
some (faint) evidence that the taste-splitting for Karsten-Wilczek fermions might be smaller than for staggered fermions.
Evidently nothing is known at this point about a potential change as a function of lattice spacing and box size, or whether it carries over to 4D and, finally, to spectroscopy.
In the event this is not a fluke, this calls for further investigation.

\bigskip

\noindent{\bf Acknowledgements}:
SD's research was supported during the initial phase of this project by the German Research Foundation DFG through SFB-TRR-55.
JHW's research was funded by DFG under project number 417533893/GRK2575 ``Rethinking Quantum Field Theory''.
This paper carries the preprint number HU-EP-22/11-RTG.



\appendix


\section{Notation and Clifford algebra conventions\label{app:notation}}


Throughout this article $\pad_\mu$ and $\pad_\mu^*$ denote the discrete forward an backward derivative, respectively,
and $\nab_\mu=(\pad_\mu+\pad_\mu^*)/2$ is the symmetric derivative.
These operators are gauged in the obvious manner; for instance the covariant symmetric derivative is
\beq
a\nab_\mu\ps(x)=\frac{1}{2}\,\Big[U_\mu(x)\ps(x+\hat\mu)-U_\mu\dag(x-\hat\mu)\ps(x-\hat\mu)\Big]
\eeq
where $U_\mu(x)$ is the parallel transporter from $x+\hat\mu$ to $x$, and $\hat\mu$ denotes $a$ times the unit-vector in direction $\mu$.
Similarly, $\lap_\mu=\pad_\mu^*\pad_\mu=\pad_\mu\pad_\mu^*$ denotes the second discrete derivative
\beq
a^2\lap_\mu\ps(x)=U_\mu(x)\ps(x+\hat\mu)-2\ps(x)+U_\mu\dag(x-\hat\mu)\ps(x-\hat\mu)
\eeq
in the presence of a gauge field $U_\mu(x)$, and one defines $aD_\mu(x,y)=a\nab_\mu(x,y)$ and
\beq
C_\mu(x,y)=\frac{1}{2}\,\Big[U_\mu(x)\de_{x+\hat\mu,y}+U_\mu\dag(x-\hat\mu)\de_{x-\hat\mu,y}\Big]
=\frac{1}{2}a^2\lap_\mu(x,y)+\de_{x,y}
\;.
\label{def_cmu}
\eeq

In $d$ Euclidean space-time dimensions ($d$ even) it is customary to use a $2^{d/2}$-dimensional representation of the $\ga$-matrices.
In $d=2$ dimensions we use $\ga_1=\si_1,\ga_2=\si_2$, along with
\beq
\gaf=-\ri\ga_1\ga_2=-\ri\si_1\si_2=\si_3=\mr{diag}(+1,-1)
\label{def_gamma5}
\eeq
and the matrices relevant to the Borici-Creutz discretization then take the form
\bea
\Gamma&=&
\frac{1}{\sqrt{2}}
\Big(\si_1+\si_2\Big)
=
\frac{1}{\sqrt{2}}\,
\bigg(\!
\begin{array}{cc}
0&1-\ri \\ 1+\ri&0
\end{array}
\!\bigg)
=
\bigg(\!
\begin{array}{cc}
0&e^{-\ri\pi/4} \\ e^{+\ri\pi/4}&0
\end{array}
\!\bigg)
\\
\si_1'&=&\Gamma\si_1\Gamma=\frac{1}{2}(\si_1+\si_2)\si_1(\si_1+\si_2)
=\frac{1}{2}(\si_1+\si_2+\si_2+\si_2\si_1\si_2)
=\si_2
\\
\si_2'&=&\Gamma\si_2\Gamma=\frac{1}{2}(\si_1+\si_2)\si_2(\si_1+\si_2)
=\frac{1}{2}(\si_1\si_2\si_1+\si_1+\si_1+\si_2)
=\si_1
\;.
\eea

Specifically for staggered and Adams fermions one defines the 1-hop operators
\bea
\Gamma_\mu(x,y)&=&\frac{1}{2}\et_\mu(x)\Big[U_\mu(x)\de_{x+\hat\mu,y}+U_\mu\dag(x-\hat\mu)\de_{x-\hat\mu,y}\Big]=\et_\mu(x)C_\mu(x)
\label{def_GA}
\\
   \Xi_\mu(x,y)&=&\frac{1}{2}\ze_\mu(x)\Big[U_\mu(x)\de_{x+\hat\mu,y}+U_\mu\dag(x-\hat\mu)\de_{x-\hat\mu,y}\Big]=\ze_\mu(x)C_\mu(x)
\label{def_XI}
\eea
with $\et_\mu(x)=(-1)^{\sum_{\nu<\mu}x_\nu}$ and $\ze_\mu(x)=(-1)^{\sum_{\nu>\mu}x_\nu}$.
Based on this we define in $d=2$ dimensions
\bea
\Gamma_5\equiv\Gamma_{50}&\equiv&-\frac{\ri}{2}[\Gamma_1,\Gamma_2]=-\frac{\ri}{2}(\Gamma_1\Gamma_2-\Gamma_2\Gamma_1)
\label{def_Gamma5}
\\
   \Xi_5\equiv\Gamma_{05}&\equiv&+\frac{\ri}{2}[   \Xi_1,   \Xi_2]=+\frac{\ri}{2}(   \Xi_1   \Xi_2-   \Xi_2   \Xi_1)
\label{def_Xi5}
\eea
where the factor in front of (\ref{def_Gamma5}) is chosen to match the one in front of (\ref{def_gamma5}).
Evidently, in $d=4$ dimensions $\Gamma_5=\Gamma_5(x,y)$ and $\Xi_5=\Xi_5(x,y)$ become 4-hop operators.
Note that both $\Gamma_5$ and $\Xi_5$ are $\ep$-hermitean operators, as follows
from (\ref{ga50_is_eps_ga05}, \ref{ga05_is_eps_ga50}) along with $\Gamma_5=\Gamma_5\dag$ and $\Xi_5=\Xi_5\dag$.

In practice all the occurences of $U_\mu(x)$ in this appendix (in $\nab_\mu$, $\lap_\mu$, $C_\mu$, $\Gamma_\mu$, $\Xi_\mu$ and thus in $\Gamma_5$ and $\Xi_5$)
are replaced by the smeared gauge field $V_\mu(x)$, in our case via one stout step \cite{Morningstar:2003gk}.


\section{Fermionic topological charges\label{app:charges}}


In the continuum one finds the formula $q_\mr{fer}[A]=(-1)^{d/2}\lim_{m\to0}m\,\mr{tr}(D_{m}^{-1}[A]\gaf)$ for the topological charge $q\in\mathbb{Z}$ of a gauge field $A_\mu(x)$.
On the lattice similar formulas hold true, provided some diligence is applied to the ``$\lim_{m\to0}$ procedure'' and the multiplicity is divided out \cite{Smit:1986fn,Smit:1987fq}.
In the following we use the sign for $d=2$; it is straightforward to adjust this for $d=4$.

\begin{figure}[!tb]
\includegraphics[width=0.49\textwidth]{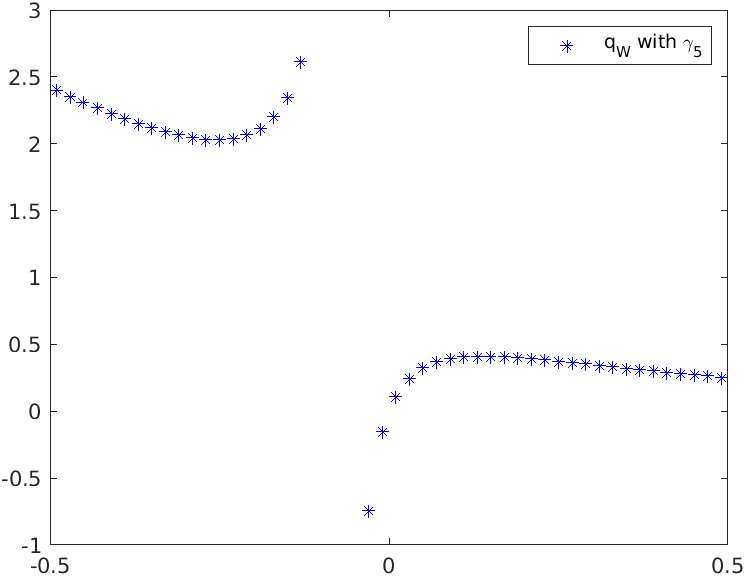}\hfill
\includegraphics[width=0.49\textwidth]{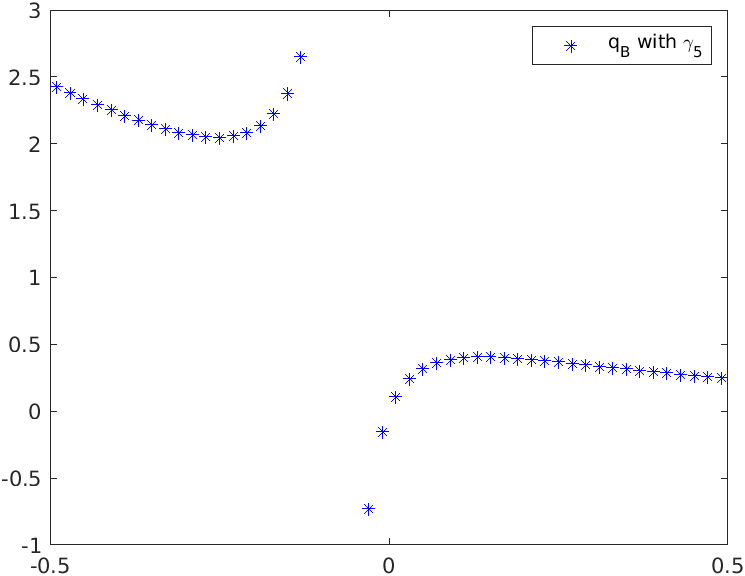}%
\vspace*{-2mm}
\caption{\label{fig:q_wils_bril}\sl
Topological charge of the Wilson (left) and Brillouin (right) operator versus $am$.}
\end{figure}

In Fig.~\ref{fig:q_wils_bril} we plot the behavior of the Wilson and Brillouin topological charges
\bea
q_\mr{W}[U]&=&-m\,\mr{tr}[(\DW+m)^{-1}I{\otimes}\gaf]
\label{def_qwils}
\\
q_\mr{B}[U]&=&-m\,\mr{tr}[(\DB+m)^{-1}I{\otimes}\gaf]
\label{def_qbril}
\eea
as a function of $m$.
Evidently, these charges are not integer-valued, and it seems there is a pole structure in the vicinity of $m\simeq-0.1$,
where the latter value coincides with the bare masses which would render $\DW+m$ or $\DB+m$ effectively massless, as seen in Figs.~\ref{fig:eigstem_wils}, \ref{fig:eigstem_bril}.

\begin{figure}[!tb]
\includegraphics[width=0.49\textwidth]{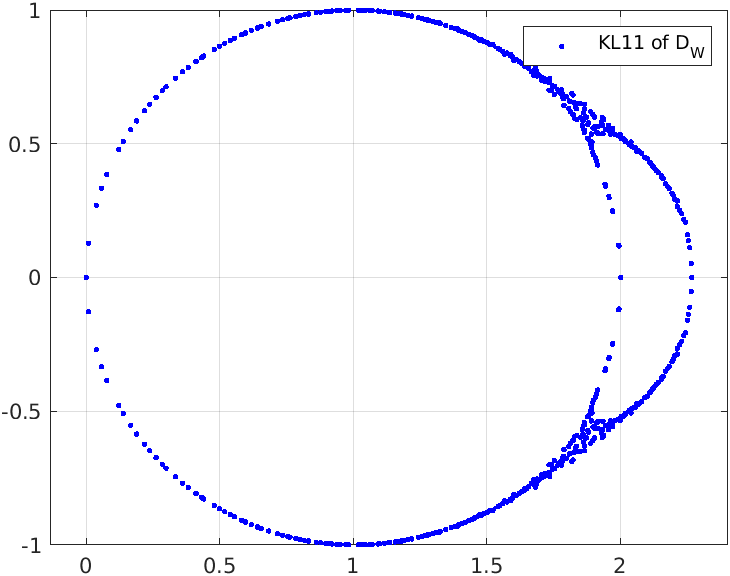}\hfill
\includegraphics[width=0.49\textwidth]{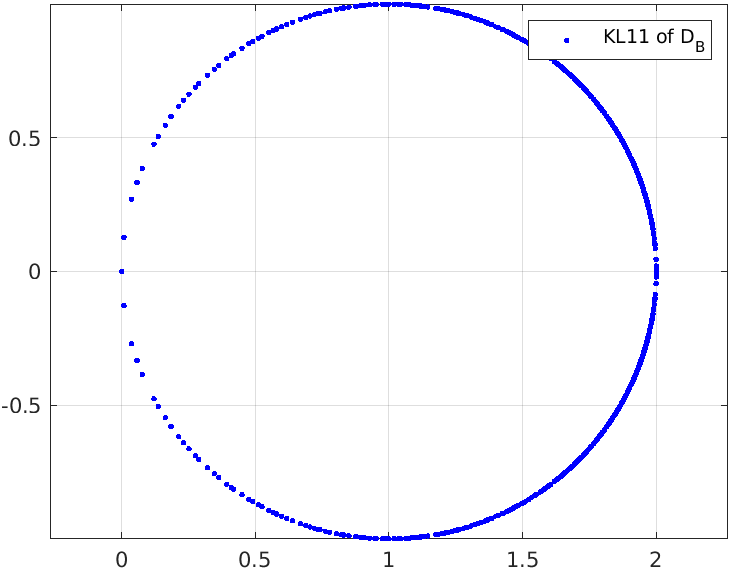}%
\\
\includegraphics[width=0.49\textwidth]{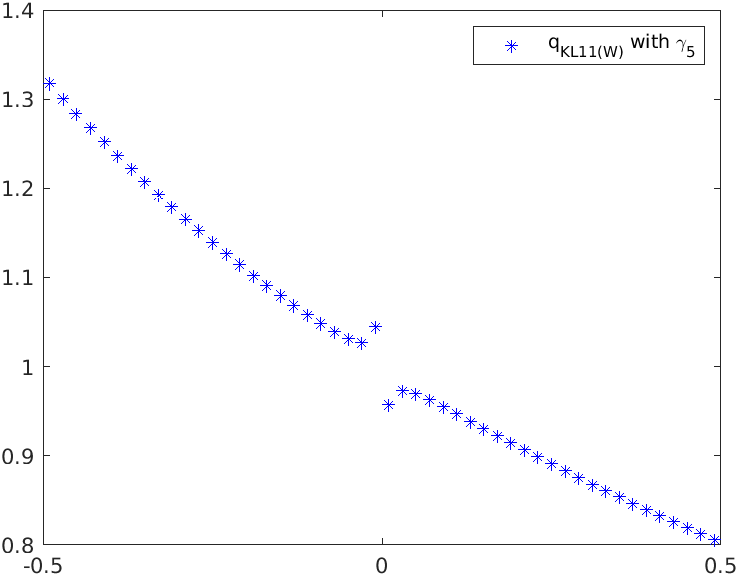}\hfill
\includegraphics[width=0.49\textwidth]{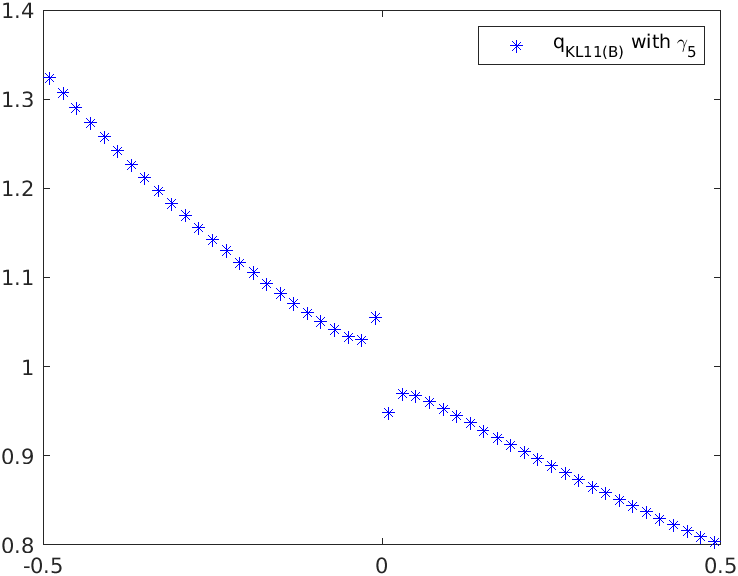}%
\vspace*{-2mm}
\caption{\label{fig:q_wilsKL11_brilKL11}\sl
Eigenvalues of the chirally improved Wilson and Brillouin operator (top), defined via one KL11 iteration (see text),
along with their topological charges versus $am$ (bottom).}
\end{figure}

Things become clearer upon considering a chirally improved version of the Wilson Dirac operator.
Using $A_\mr{W}\equiv(\DW-1)\dag(\DW-1)$, a chirally improved descendant at zero mass is defined via one KL11 iteration as $\DW^\mr{KL11}=(\DW-1)\frac{A_\mr{W}+3}{3A_\mr{W}+1}+1$,
and $\DB^\mr{KL11}$ is defined similarly, see Ref.~\cite{Durr:2017wfi} for details.
The eigenvalue spectra of these operators are shown in the top panels of Fig.~\ref{fig:q_wilsKL11_brilKL11}.
One notices that these operators show very small additive mass renormalization,
and a few more Kenney-Laub steps make it zero within machine precision \cite{Durr:2017wfi}.
Upon plugging $\DW^\mr{KL11}$ in place of $\DW$ into (\ref{def_qwils}) and $\DB^\mr{KL11}$ in place of $\DB$ into (\ref{def_qbril}),
one obtains the results in the bottom panels of Fig.~\ref{fig:q_wilsKL11_brilKL11}.
This time it is clear that one should ignore a small ``pole area'' near $m=0$, and read off the curve at a nearby $m$-value to find $q_\mr{W}=1$ and $q_\mr{B}=1$.

\begin{figure}[!tb]
\includegraphics[width=0.49\textwidth]{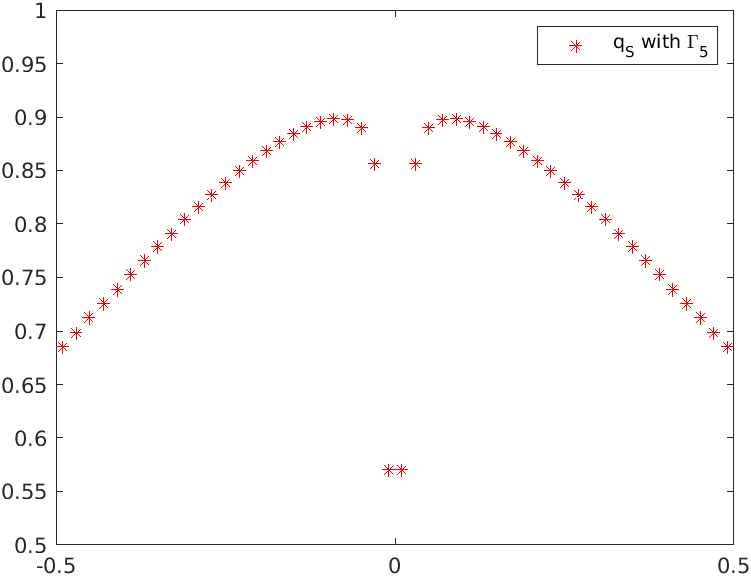}\hfill
\includegraphics[width=0.49\textwidth]{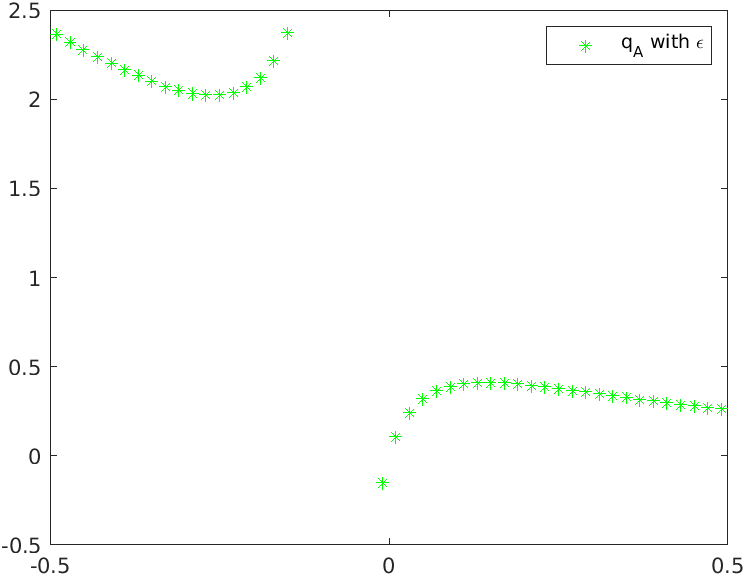}%
\vspace*{-2mm}
\caption{\label{fig:q_stag_adam}\sl
Topological charge of the staggered (left) and Adams (right) operators versus $am$,
using $\Gamma_5=\Gamma_{50}$ as chirally sensitive probe in the former case, and $\ep=\Gamma_{55}$ in the latter case.}
\end{figure}

Given the arguments presented in Sec.~\ref{sec:stagadam} the staggered and Adams charges are
\bea
q_\mr{S}[U]&=&-\frac{m}{2}\,\mr{tr}[(\DS              +m)^{-1}\Gamma_{50}]
\\
q_\mr{A}[U]&=&-          m\,\mr{tr}[(\DS+1-\Gamma_{05}+m)^{-1}\Gamma_{55}]
\eea
on a background $U$; they are shown as a function of $m$ in Fig.~\ref{fig:q_stag_adam}.
In the former case the function is even in $m$, in the latter case the behavior is similar to that of $q_\mr{W}$ and $q_\mr{B}$.
Again, one should ignore a ``pole area'' near $m=0$, and read off the curve at a nearby $m$-value to find $q_\mr{S}=1$ and $q_\mr{A}=1$ (after some suitable interpolation and renormalization).

\begin{figure}[!tb]
\centering
\includegraphics[width=0.49\textwidth]{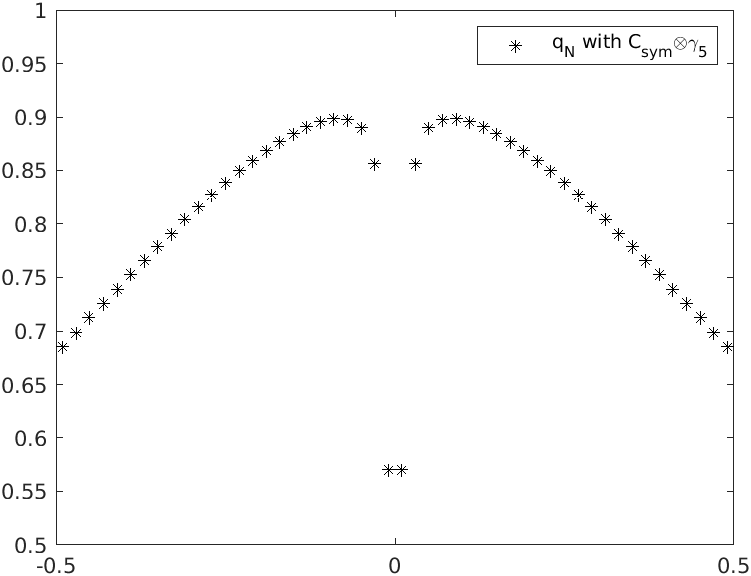}%
\includegraphics[width=0.49\textwidth]{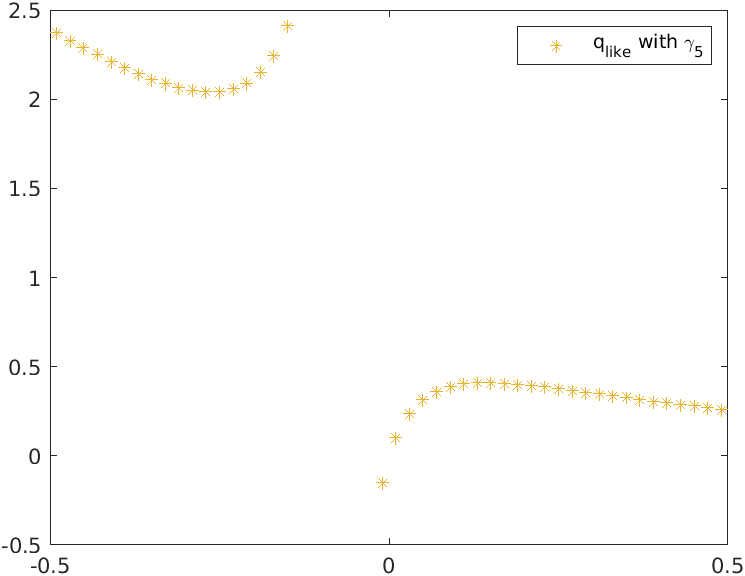}%
\vspace*{-2mm}
\caption{\label{fig:q_naiv}\sl
Topological charge of the operator $\DN+m$ with $C_\mr{sym}\otimes\gaf$ as probe versus $am$ (left),
and of $D_\mr{like}+m=\DN+1-C_\mr{sym}+m$ with $\gaf$ as probe versus $am$ (right).}
\end{figure}

From the arguments presented in Sec.~\ref{sec:naiv} it follows that one should define the charges
\bea
q_\mr{N}   [U]&=&-\frac{m}{4}\,\mr{tr}[(\DN             +m)^{-1}C_\mr{sym}{\otimes}\gaf]
\\
q_\mr{like}[U]&=&-\frac{m}{2}\,\mr{tr}[(\DN+1-C_\mr{sym}+m)^{-1}         I{\otimes}\gaf]
\eea
for the naive and Adams-like operator, respectively.
Their dependence on $m$ is displayed in Fig.~\ref{fig:q_naiv}.
Again, the former charge is even in $m$, the latter one is not.

\begin{figure}[!tb]
\includegraphics[width=0.49\textwidth]{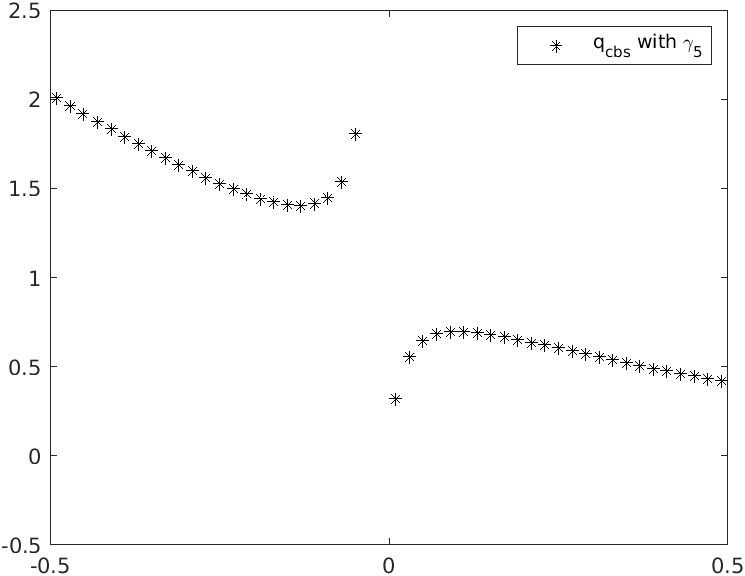}\hfill
\includegraphics[width=0.49\textwidth]{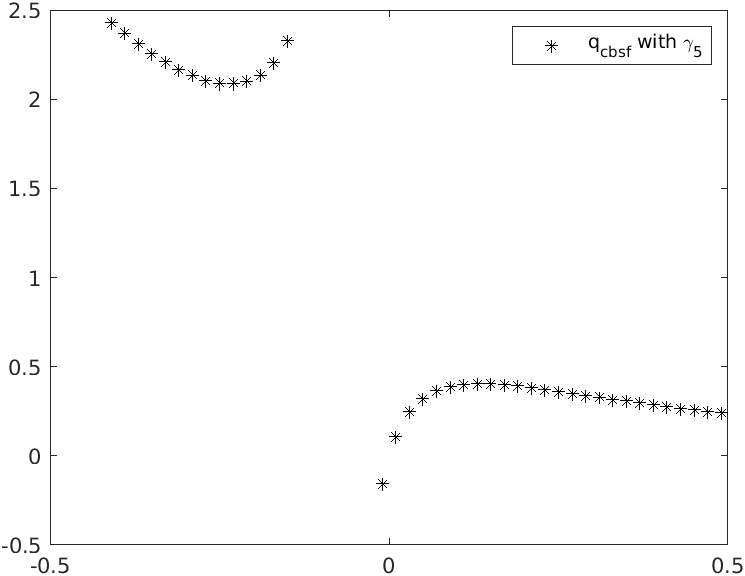}%
\vspace*{-2mm}
\caption{\label{fig:q_cbsq_cbsf}\sl
Topological charge of the ``central-branch-squared'' operator at $r=1$ (left) and ``central-branch-squared-and-flipped'' operator at $r=\frac{1}{4}$ (right) versus $am$.}
\end{figure}

Given the arguments presented in Sec.~\ref{sec:cebr} it is clear that the natural definitions are
\bea
q_\mr{cbs} [U]&=&+\frac{m}{2}\,\mr{tr}[(D_\mr{cbs} +m)^{-1} I{\otimes}\gaf]
\\
q_\mr{cbsf}[U]&=&-\frac{m}{2}\,\mr{tr}[(D_\mr{cbsf}+m)^{-1} I{\otimes}\gaf]
\eea
for ``central-branch-squared'' and ``central-branch-squared-and-flipped'' fermions, respectively.
The respective plots are found in Fig.~\ref{fig:q_cbsq_cbsf}, based on $r=1$ in the former and $r=\frac{1}{4}$ in the latter case.
The former formulation benefits from a rather small additive mass renormalization.

\begin{figure}[!tb]
\includegraphics[width=0.49\textwidth]{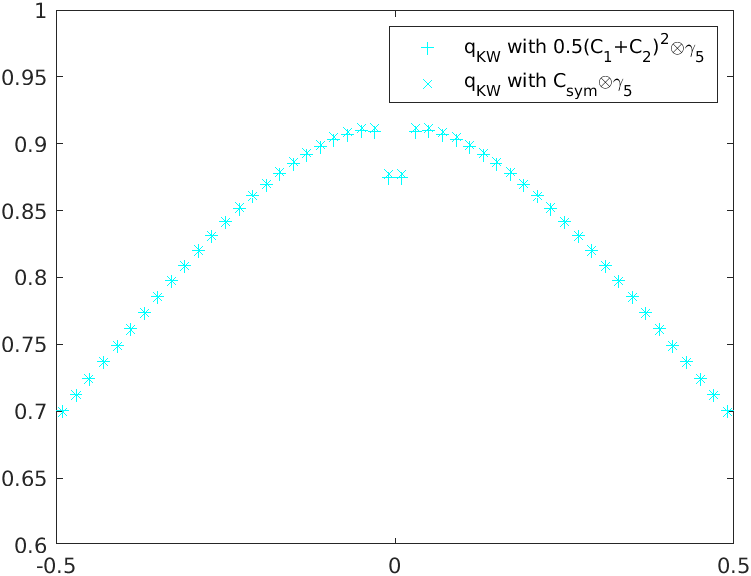}\hfill
\includegraphics[width=0.49\textwidth]{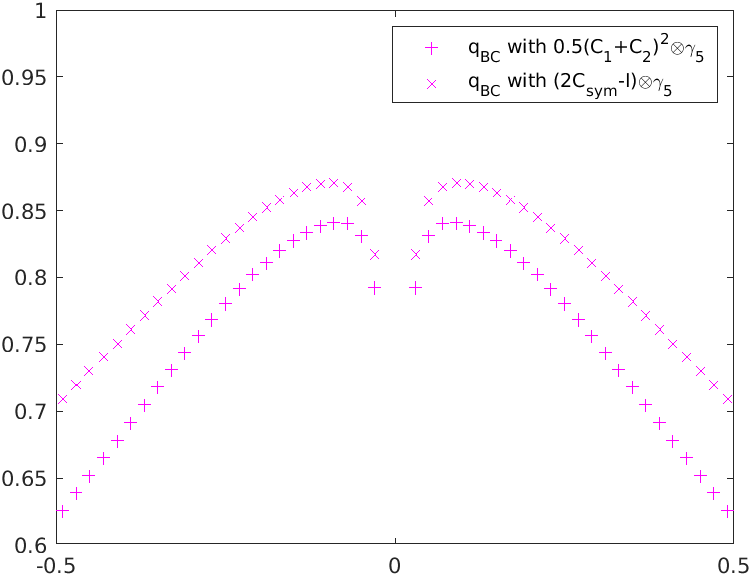}%
\vspace*{-2mm}
\caption{\label{fig:q_kawi_bocr}\sl
Topological charge of the KW (left) and BC (right) operator versus $am$.
In either case a chirality operator based on $\frac{1}{2}(C_1+C_2)^2$ and one based on $C_\mr{sym}$ are used.}
\end{figure}

Finally, following the discussion in Secs.~\ref{sec:kawi} and \ref{sec:bocr}, we define the charges
\bea
q_\mr{KW}[U]&=&-\frac{m}{2}\,\mr{tr}[(\DKW+m)^{-1}\frac{1}{2}(C_1+C_2)^2{\otimes}\gaf]
\\
q_\mr{KW}[U]&=&-\frac{m}{2}\,\mr{tr}[(\DKW+m)^{-1}            C_\mr{sym}{\otimes}\gaf]
\\
q_\mr{BC}[U]&=&-\frac{m}{2}\,\mr{tr}[(\DBC+m)^{-1}\frac{1}{2}(C_1+C_2)^2{\otimes}\gaf]
\\
q_\mr{BC}[U]&=&-\frac{m}{2}\,\mr{tr}[(\DBC+m)^{-1}       [2C_\mr{sym}-1]{\otimes}\gaf]
\eea
for Karsten-Wilczek and Borici-Creutz fermions, respectively.
The four curves%
\footnote{We checked that using $X=[\frac{1}{2}(C_1+C_2)^2-1]{\otimes}\gaf$ instead of $X=\frac{1}{2}(C_1+C_2)^2{\otimes}\gaf$ brings no visible change in either panel of Fig.~\ref{fig:q_kawi_bocr}.
This is in line with our statements regarding the ``needle plots'' (or diagonal part of the chirality operator) in Sec.~\ref{sec:kawi} and Sec.~\ref{sec:bocr}.}
are displayed in Fig.~\ref{fig:q_kawi_bocr}; they are even in $m$.
For $\DKW$ the two options of the charge operator work equally well, in line with what we reported in Sec.~\ref{sec:kawi}.
For $\DBC$ the chirality operator $[2C_\mr{sym}-1]{\otimes}\gaf$ seems to work marginally better than $\frac{1}{2}(C_1+C_2)^2{\otimes}\gaf$, again in line with Sec.~\ref{sec:bocr}.

In summary, a fermionic topological charge can be determined by reading off the $m$-dependent charge ``slightly to the left'' and ``slightly to the right'' of the pole-like structure,
and applying some suitable average (and possibly some renormalization and a cast-to-integer operation).
The averaging procedure could be formalized, but it is clear that some arbitrariness remains.
In practice all occurrences of $U_\mu(x)$ in this appendix are replaced by the smeared gauge field $V_\mu(x)$.
This holds for all Dirac matrices $D$ and the staggered $\Gamma_{50}$, $\Gamma_{05}$, in line with App.~\ref{app:notation}.
Last but not least, in 4D the multiplicity factors need to be adjusted.
In the staggered case it is $\frac{1}{4}$, in the Adams case $\frac{1}{2}$, the naive operator has $\frac{1}{16}$, and Adams-like operator $\frac{1}{8}$.
For $D_\mr{cbs}$ it is $\frac{1}{6}$, and for $D_\mr{cbsf}$ nothing changes.


\section{Analytic argument\label{app:analytic}}


It is not surprising that the continuum formula $q_\mr{fer}[A]=(-1)^{d/2}\lim_{m\to0}m\,\mr{tr}(D_{m}^{-1}[A]\gaf)$
has lattice counterparts as discussed in App.~\ref{app:charges}.
In the following we omit the factor $(-1)^{d/2}$ and concentrate on the Wilson operator $\DW$, but we see no obstacle to applying the argument to any other action.
The argument is not entirely new \cite{Smit:1986fn,Smit:1987fq}, but it is still elucidating.

We want to feed the trace formula $q[U]=m\,\mr{tr}(D_{m}^{-1}[U]\gaf)$ with the mode representation $D_{m}=\sum_i(\la_i+m)|\ps_i\>\<\ps_i|$ of the Dirac operator,
where $|\ps_i\>$ is the right-eigenvector of $D$ and $\<\ps_i|$ is the left-eigenvector (cf.\ footnote \ref{foot:normality}) on the gauge background $U$.
In this representation the inverse is given by $D_{m}^{-1}=\sum_i(\la_i+m)^{-1}|\ps_i\>\<\ps_i|$, thanks to the bi-orthogonality condition $\<\ps_i|\ps_j\>=\de_{ij}$.
This yields $q=m\,\mr{tr}(\sum_i(\la_i+m)^{-1}|\ps_i\>\<\ps_i|\gaf)=m\sum_i(\la_i+m)^{-1}\mr{tr}(|\ps_i\>\<\ps_i|\gaf)$.
Due to the cyclic property of the trace the last factor is $\mr{tr}(\<\ps_i|\gaf|\ps_i\>)$,
and since the trace of a scalar object is just that object we have $q=m\sum_i(\la_i+m)^{-1}\<\ps_i|\gaf|\ps_i\>$.

Next we should recall the ``needle plots'' for each action, for instance the right panel in Fig.~\ref{fig:eigstem_wils} in case of $\DW$.
Since $\<\ps_i|\gaf|\ps_i\>\simeq0$ for all $i$ with $\mr{Im}(\la_i)$ significantly non-zero,
only the would-be zero-modes in the physical branch and their siblings in the lifted branches contribute to this sum.
But for these chiral modes $\<\ps_i|\gaf|\ps_i\>\simeq\pm1$, hence we have
\beq
q_\mr{lat}=m\sum_{i\in\mr{needles}}(\la_i+m)^{-1}\si_i
\label{sumoverneedles}
\eeq
where $\si_i$ is the sign of the needle associated with the exactly real mode $i$.

At this point we need to distinguish the various formulations and the dimensionality of space-time.
For instance for $\DW$ in $d=2$ dimensions we have three contributions
\beq
q_\mr{W}=m\,\big\{ \frac{1}{\la_0+m} - \frac{2}{\la_\mr{1-lift}+m} + \frac{1}{\la_\mr{2-lift}+m} \big\}
\label{alternatingsigncontributions}
\eeq
with alternating signs and weights reflecting the multiplicity of each branch.
And for $\DW$ in $d=4$ dimensions the sign and weight sequence would be $\{+1,-4,+6,-4,+1\}$.

Next one should  take into account that the two downward pointing needles in Fig.~\ref{fig:eigstem_wils} are related by reflection symmetry.
In fact the entire eigenvalue spectrum of the massless operator is symmetric about $\mr{Re}(\la)=dr$, and for the massive operator this vertical reflection line is at $\mr{Re}(\la)=dr+m$.
Moreover, the eigenvalues in (\ref{alternatingsigncontributions}) are exactly real, and the offset of $\la_0$ is basically given by the additive mass shift.
Hence $\la_0\simeq-m_\mr{crit}$, where $m_\mr{crit}<0$ denotes the bare mass $m$ at which $\DN$ creates massless pions (on an infinite lattice).
In consequence
\beq
q_\mr{W}=m\,\big\{ \frac{1}{m-m_\mr{crit}} - \frac{2}{2r+m} + \frac{1}{4r+m_\mr{crit}+m} \big\}
\eeq
where we take into account that $\la_\mr{1-lift}=2r$ for $m=0$, modulo the ``horizontal fuzziness'' discussed in Sec.~\ref{sec:cebr}.
Bringing everything atop a common denominator yields
\beq
q_\mr{W}=m
\frac{ (2r+m_\mr{crit})^2 + (2r+m)^2}{ (m-m_\mr{crit}) (2r+m) (4r+m_\mr{crit}+m) }
\eeq
and for $m\simeq m_\mr{crit}$ the latter expression simplifies to
\beq
q_\mr{W}\simeq m
\frac{ 2(2r+m_\mr{crit})^2 }{ (m-m_\mr{crit}) (2r+m_\mr{crit}) (4r+2m_\mr{crit}) }
=\frac{m}{m-m_\mr{crit}}
\label{polestructure}
\eeq
which underpins the pole structure that shows up near $m_\mr{crit}<0$ in the right panel of Fig.~\ref{fig:q_wils_bril}.
In fact, for the chirally improved Wilson operator $\DW^\mr{KL11}$ the lower left panel in Fig.~\ref{fig:q_wilsKL11_brilKL11}
demonstrates that its additive mass shift is much smaller than that of the Wilson operator.

For doubled chiral actions the basic formula (\ref{sumoverneedles}) still holds true, but the sum is dominated by the would-be zero-modes which come in complex conjugate pairs.
Here it is important that the $\si_i$ reflect the chiralities as determined by an appropriate chirality operator.
For instance, for staggered fermions $\si_i$ refers to $\Gamma_5$ so that the pairs have \emph{like sign} (with $\ep$ they would have opposite sign and thus cancel).
Specifically in $d=2$ dimensions one has
\beq
q_\mr{S}=\frac{m}{2}\,\big\{ \frac{1}{+\ri\ep+m} + \frac{1}{-\ri\ep+m} \big\}
=\frac{m}{2}\,\frac{2m}{(\ri\ep+m)(-\ri\ep+m)}=\frac{m^2}{\ep^2+m^2}
\eeq
which suggests that there is a double-pole structure near $m=0$.
In $d=4$ dimensions the factor in front is $\frac{m}{4}$ and there are $4|q|$ contributions, so the conclusion is unchanged.
Similarly, the argument goes through for KW and BC fermions.

Looking at the plots assembled in App.~\ref{app:charges} we find the prediction of the pole structure confirmed, both for non-chiral and chiral (doubled) actions.
However, it is clear that the curves include a significant regular part which is not covered by the argument.
Regarding the non-chiral actions we comment that mapping out the pole structure in (\ref{polestructure}) provides a handle at $\la_0$
(on a given configuration) and thus at the additive mass shift $-m_\mr{crit}$ (after averaging over configurations).
This approach does not require any eigenvalue and/or eigenvector computation, nor does it involve spectroscopy, but we do not know whether it is very practical.








\clearpage



\end{document}